\documentclass[%
 reprint,
 superscriptaddress,
 amsmath,amssymb,
 aps,
 pra,
]{revtex4-1}

\usepackage{graphicx}% Include figure files
\usepackage{dcolumn}% Align table columns on decimal point
\usepackage{bm}% bold math
\usepackage{hyperref}% add hypertext capabilities
\usepackage{longtable}
\usepackage{float}
\usepackage[mathlines]{lineno}% Enable numbering of text and display math
\usepackage{hyperref}
\hypersetup{
    colorlinks=true,
    linkcolor=blue,
    urlcolor=blue,  
}

\usepackage{xcolor}
\usepackage{amsfonts, amsmath, amssymb, amsthm}
\usepackage{braket}

\graphicspath{ {./figures/} }

\begin{document}

\preprint{APS/123-QED}

\title{Advancing Free-Space Optical Communication System Architecture:\\
Performance Analysis of Varied Optical Ground Station Network Configurations}

\author{Eugene Rotherham}
\affiliation{BryceTech, United Kingdom}
\thanks{These authors contributed equally to this work.}
\author{Connor Casey}
\email{cacasey@umass.edu}
\affiliation{College of Information and Computer Sciences, University of Massachusetts Amherst, Amherst, MA, USA}
\thanks{These authors contributed equally to this work.}
\affiliation{Department of Physics, University of Massachusetts Amherst, Amherst, MA, USA}
\thanks{These authors contributed equally to this work.}
\author{Eva Fernandez Rodriguez}
\affiliation{Netherlands Organisation for Applied Scientific Research (TNO), The Netherlands}
\author{Karen Wendy Vidaurre Torrez}
\affiliation{Kyushu Institute of Technology, Japan}
\author{Maren Mashor}
\affiliation{National Space Research and Development Agency (NASRDA), Nigeria}
\author{Isaac Pike}
\affiliation{University College London (UCL), United Kingdom}

\date{October 3, 2024}

\begin{abstract}
Free-Space Optical (FSO) communication architectures are increasingly being adopted as an alternative to traditional radio-frequency methods on modern space-based systems, e.g., small-satellite mega-constellations (SpaceX’s Starlink), data-relay (European Data Relay System), and deep-space communications (NASA’s Psyche). The demand for higher data transmission rates, quantum cryptography techniques, license-free operation, and smaller and more cost-effective terminals has led to significant commercialization of FSO communication technology within the space sector. Currently, the space-to-space optical link market segment presents opportunities orders of magnitude higher than the space-to-ground segment due to the limited technological maturity and business viability factors of optical data transmission to Earth. The projected increase in laser communication terminals aboard satellites over the next decade presents opportunities to develop the ground segment and highlights the urgent need for agile space-ground infrastructure design to meet the ever-evolving data demands effectively.

Next-generation space-to-ground optical communication networks must be designed to maximize overall data throughput and system availability while remaining affordable to procure and operate. Availability of FSO systems is predominantly influenced by localized cloud cover, whereas link performance is affected by atmospheric turbulence and respective Optical Ground Station (OGS) characteristics, such as transmission power and tracking accuracy. Currently, high-capacity large OGS require significant capital investment and resources to build and operate. A global network of smaller portable OGS is envisaged to fulfill the role of a few large OGS by providing system flexibility and resilience through high spatial diversity and terminal redundancy at a fraction of the cost, while compromising on individual terminal link performance.

This study discusses the current state of FSO technology, as well as global trends and developments in the industrial ecosystem to identify obstacles to the full realization of optical space-to-ground communication networks. Additionally, link performance and network availability trade-off studies are presented, comparing overall system performance between portable and large OGS networks in conjunction with a constellation of small low Earth orbit (LEO) satellites. The paper provides an up-to-date overview and critical analysis of the FSO industry and assesses the feasibility of low-cost portable terminals as an alternative to larger high-capacity OGS systems. This initiative aims to better inform optical communications stakeholders, including governments, academic institutions, satellite operators, manufacturers, and communication service providers.
\end{abstract}

\maketitle

\section{Introduction}\label{chap:introduction}

    Free-space Optical Communications (FSOC) have garnered significant attention in recent years, not only for their numerous applications in broadband communications and Earth Observation (EO) but also for their suitability in Quantum Key Distribution (QKD). For broadband communications, optical frequencies can be used on feeder links for both Geostationary Earth Orbit (GEO) and Non-GEO (NGSO) communication systems \cite{kaushal_2017, Lyras2019, Toyoshima2005}. This is evident in the plethora of developments involving FSO technology ongoing in the space sector, such as the provision of small-satellite mega-constellations, data-relay systems, and deep-space communications. The growing demand for higher data transmission rates, quantum cryptography techniques, licence-free operation, and smaller, more cost-effective terminals has driven the significant commercialisation of FSO technology within the space sector. At present, the space-to-space optical link market offers opportunities that are vastly greater than the space-to-ground segment, due to the latter's limited technological maturity and business feasibility. However, the anticipated increase in laser communication terminals on satellites over the next decade presents opportunities for ground segment development, underlining the urgent need for flexible space-ground infrastructure designs to meet evolving data demands.

    \begin{figure}[]  \includegraphics[width=\linewidth]{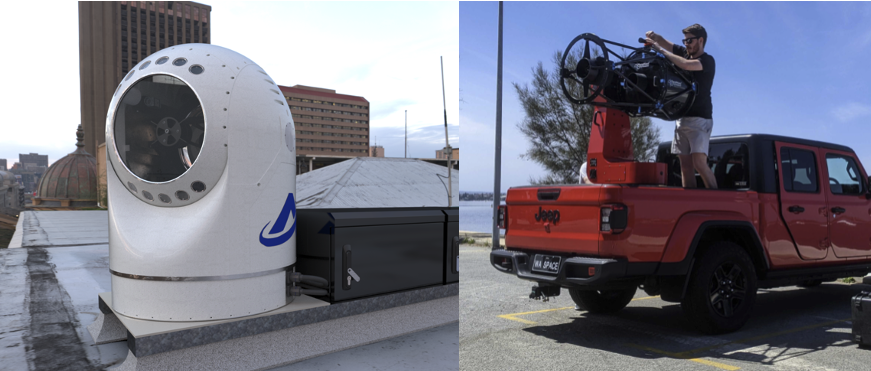} \caption{ \textbf{Real-World Examples of Mobile OGS} [Left] Archangel Lightworks Terra-M small Optical Ground Stations (OGS) utilised for establishing ground-based communication links. [Right] TeraNet-3 mobile OGS terminal designed for flexible deployment and on-site optical signal processing \cite{archangel_2024, rattenbury_2024}. } \label{fig:smallogs} \end{figure}

        Free-space Optical Communications (FSOC) have garnered significant attention in recent years, not only for their numerous applications in broadband communications and Earth Observation (EO) but also for their suitability in Quantum Key Distribution (QKD). For broadband communications, optical frequencies can be used on feeder links for both Geostationary Earth Orbit (GEO) and Non-GEO (NGSO) communication systems \cite{kaushal_2017, Lyras2019, Toyoshima2005}. This is evident in the plethora of developments involving FSO technology ongoing in the space sector, such as the provision of small-satellite mega-constellations, data-relay systems, and deep-space communications. The growing demand for higher data transmission rates, quantum cryptography techniques, licence-free operation, and smaller, more cost-effective terminals has driven the significant commercialisation of FSO technology within the space sector. At present, the space-to-space optical link market offers opportunities that are vastly greater than the space-to-ground segment, due to the latter's limited technological maturity and business feasibility. However, the anticipated increase in laser communication terminals on satellites over the next decade presents opportunities for ground segment development, underlining the urgent need for flexible space-ground infrastructure designs to meet evolving data demands.
    
    Commercially available small OGS (aperture < 0.5m), particularly mobile units as shown in Figure \ref{fig:smallogs}, have significantly reduced upfront and operational costs when compared to larger stationary OGS systems. Furthermore, they can be rapidly deployed in highly spatially distributed network configurations, providing benefits such as increased availability and system redundancy \cite{riesing_2018}. Mobile OGS have already been integrated into several existing OGS networks, such as the Australian Optical Ground Station Network (AOGSN) and Chinese QKD experiments \cite{rattenbury_2024, Lu_2022}. The proliferation of commercially available small and mobile OGS has made their procurement and deployment easier than ever.

    However, the performance of smaller OGS systems faces challenges, such as more difficult alignment and tracking, limited computational power, and smaller aperture sizes, which affect sensitivity and data throughput \cite{riesing_2018}. Despite these challenges, advancements in optics have enabled smaller ground stations to be used effectively in high-capacity network settings, showing promising results. To facilitate the widespread adoption of small or portable OGS, it is crucial to quantitatively assess their overall performance compared to larger legacy systems, thereby evaluating their true added value to the FSOC landscape.

    This research aims to provide evidence-based insights and recommendations to FSO stakeholders through critical industry analysis and network simulations. By quantifying the trade-offs associated with integrating small optical ground stations into existing or new communication networks, stakeholders can assess how these smaller stations might reshape the future communications landscape.

    The first section of this paper examines the current state of FSO technology, along with global trends and developments in the industrial ecosystem. It identifies the obstacles hindering the complete realisation of optical space-to-ground communication networks. This section also provides an up-to-date overview and critical analysis of the FSO industry, assessing the feasibility of low-cost portable terminals as alternatives or supporting to larger high-capacity OGS networks.

    The second part discusses a method for modelling laser communication network performance. Preliminary results are presented, comparing the link performance and network availability for EO downlinks. This comparison involves one large OGS and a supporting network of up to six smaller mobile OGS distributed around Europe, in conjunction with a small LEO remote sensing satellite.
\section{Space-Based Free-Space Optical Communications: Advantages, Challenges, and Mitigation Strategies} \label{chap: tradeoffs}

In this section, we provide a thorough evaluation of FSOC in satellite communications, exploring the key advantages and challenges of the technology compared to RF systems, and discussing current mitigation strategies employed to further improve FSOC performance and reliability.

\subsection{Key Advantages of Free-Space Optics Compared with Radio Frequency}

Satellite-based FSOC offers numerous advantages over Radio Frequency (RF) satellite systems, along with key considerations. FSO can achieve substantially higher throughput, with potential data rates ranging from gigabits per second (Gbps) to terabits per second (Tbps) in a single channel, significantly surpassing the kilobits per second (Kbps) to Gbps range typical of RF systems \cite{alkholidi_2014}. Furthermore, FSO systems enhance signal security due to their narrow beam divergence and immunity to electromagnetic interference, ensuring high reliability in environments susceptible to electronic warfare or other electromagnetic disturbances, while also maintaining a low probability of detection \cite{alkholidi_2014} and interception. In contrast, RF antennas produce side lobes, which can result in interference and security vulnerabilities. Additionally, FSO technology enables QKD, a security method that utilises quantum entanglement or photon polarisation to transmit encryption keys with near-absolute security— a capability unattainable with RF systems. Currently, FSO does not require frequency licensing, thereby eliminating the regulatory costs and complexities associated with satellite-based RF communications, such as bandwidth congestion, simplifying deployment, and reducing operational costs \cite{alkholidi_2014}.

\subsection{Key Challenges of Free-Space Optics Compared with Radio Frequency}

Optical satellite communication systems, despite offering significant advantages, encounter substantial challenges compared to RF systems for space-to-ground communication, especially due to weather and atmospheric conditions. Optical systems are highly susceptible to weather effects, with cloud cover being a primary cause of signal attenuation \cite{kaushal_2017, fuchs_2015, giggenbach_2015}. FSO beams experience degradation from particles such as aerosols and dust, as well as atmospheric gases like CO$_2$ and H$_2$O, which induce absorption and scattering \cite{atmospheric_maharjan_2022}. Atmospheric turbulence, characterised by the refractive index structure parameter $(C_n^2)$, varies with altitude, affecting uplink and downlink differently and causing beam wander that increases Bit Error Rates (BER) in the uplink \cite{atmospheric_maharjan_2022}. Additionally, FSO systems demand significantly higher pointing accuracy than RF systems, often requiring gimbals for coarse pointing with angular resolutions in the \(\mu\mathrm{rad}\) range. The costs and availability of materials, manufacturing processes, and specialised equipment for optical components can be considerable, and high-grade components are expensive \cite{optics-org}. Achieving the necessary Technology Readiness Level (TRL) for optical components necessitates extensive ground testing and validation, which can be costly. Components such as fine precision tracking must be proven to operate under various conditions, and access to satellites for testing is expensive and limited due to the scarcity of satellites equipped with space-ground optical communication capabilities \cite{fso_flight_worthiness}.

\subsection{Mitigation Strategies for Free-Space Optics}

Characterising atmospheric effects on optical links in real-time enables the forecasting of link outages \cite{csdss_green_2017}. Outages caused by clouds are often modelled as an on/off channel based on the presence of clouds, which requires the OGS operator to determine whether clouds are present along the slant path. Therefore, it is important to have long-term measurements, such as five years, with whole-sky images for each site of interest. Satellite imagery can serve as a source for historical or even real-time weather data and provide information about clouds over a single pixel—typically representing a large spatial area, for example, 3 km × 3 km—using metrics such as cloud mask or cloud coverage \cite{csdss_magenta_2022}. In \cite{csdss_green_2017}, network availability is calculated using atmospheric monitoring data while considering various cloud coverage thresholds. The findings of \cite{lyras_2020} indicate that network availability is highly sensitive to the threshold, thereby making the accuracy and resolution of cloud cover metrics significant.

Operational mitigation strategies include operating within specific wavelength windows (850 nm, 1060 nm, 1250 nm, and 1550 nm) where attenuation can be less than 0.2 dB/km \cite{csdss_magenta_2022}. Most commercially available FSO systems now utilise these wavelength windows \cite{khalighi_2014}.

Atmospheric turbulence poses significant challenges to optical communication but can be mitigated through adaptive optics techniques \cite{kalati_2019}. These methods employ wavefront sensors to measure distortions, wavefront correctors to compensate for them, and control systems to determine and apply the necessary corrections. Additionally, adaptive micro-electro-mechanical beam control techniques dynamically adjust beam size based on optical channel conditions \cite{jahid_2022}. Adjusting the receiver's diameter helps mitigate intensity fluctuations and degradation in received light caused by atmospheric turbulence. Studies have also explored auto-alignment techniques using tracking algorithms and agile receivers, with mechanical gimbals assisting in fine alignment to reduce pointing losses \cite{cap_2008}.

A hybrid configuration combining FSO and RF systems enhances communication capabilities and flexibility, allowing a switch to RF when the optical link is severely affected by fading, thus ensuring uninterrupted communication.

OGS are often located at or near astronomical telescope sites with favorable environments to minimize link outages due to cloud coverage. High-altitude sites reduce the space-to-ground link distance and limit interference from natural or man-made structures while facilitating integration with terrestrial network infrastructure. Optical gateway locations are also strategically chosen where existing fiber-optic cables and ground stations exist, such as teleports with weather monitoring infrastructure and fiber-optic network connections.

Spatial diversity of ground station sites significantly improves network availability by reducing the probability that all OGS are simultaneously affected by clouds when multiple interconnected stations are installed \cite{csdss_magenta_2022}. Several studies have investigated the optimal number of OGS for various satellite communication systems \cite{fuchs_2015, giggenbach_2015, lyras_2020}. Smart Gateway Diversity enhances this approach by enabling station switching based on real-time weather categorization, allowing accurate outage predictions and rerouting connections to available stations.

Coherent modulation techniques can significantly enhance data rates. While On-Off Keying (OOK) is the simplest and helps mitigate scintillation, Differential Phase Shift Keying (DPSK) offers better sensitivity and tolerance to fading, doubling spectral efficiency compared to OOK \cite{csdss_magenta_2022}. Pulse Position Modulation (PPM) is also utilized in many commercial components operating at high data rates up to 10 Gbps \cite{motlagh_2008, mishchenko_2015}.

Forward error correction codes like Reed-Solomon coding are commonly used in satellite telecommunications and are effective at detecting and correcting errors under moderate to strong atmospheric turbulence, where the Bit Error Rate (BER) increases. For instance, applying Reed-Solomon coding with 50\% redundancy can achieve an 6 dB improvement in Signal-to-Noise Ratio (SNR) over conventional non-error-corrected FSOC \cite{Geisel1990,Ajewole2019}.

Industry advancements in cost reduction, manufacturing techniques, and quality control for critical optical components, such as lenses, have significantly enhanced the performance and reliability of FSOC systems. This progress is further supported by research and development efforts exploring alternatives like plastic photonics \cite{plastics}.

\section{Current Satellite Free-Space Optical Communications Industrial Ecosystem}
\label{chap: ecosystem}

\begin{figure}[]
        \includegraphics[width=\linewidth]{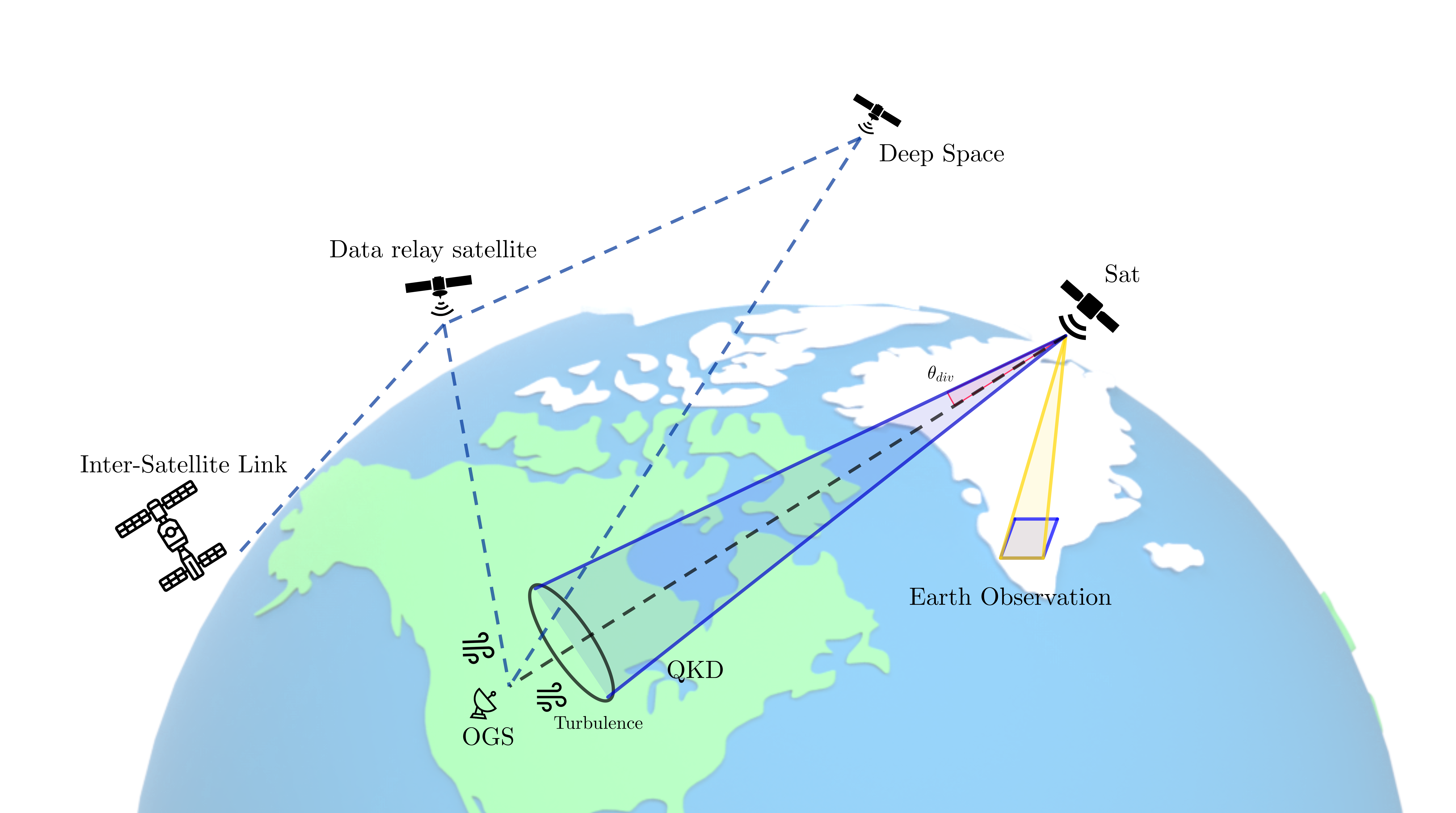}
        \caption{
        \textbf{Overview of satellite-based FSO communication applications.} The diagram shows key applications of satellite FSOC highlighting the role of data relay satellites, OGS, and atmospheric turbulence effects.
        }
        \label{fig:apps}
    \end{figure}

        This section provides an overview of the satellite-based FSOC industrial ecosystem as of 2024, summarizing key technological developments and initiatives. The key applications of space-to-ground FSOC are outlined, classified into EO, Satellite Communications (SatCom), and QKD, as illustrated in Figure \ref{fig:apps}, and provide examples of notable developments from commercial and governmental entities. Section \ref{sec: ground} breaks down the key players and developments in the OGS industry. The data were collected using a combination of primary sources, including official reports and company documents, alongside up-to-date information from reputable online news articles.

    \subsection{Earth Observation}

        Remote-sensing satellites continuously generate data from their instruments, which are transmitted via a one-way downlink to ground stations for collection and processing. The volume of data transmitted to Earth is constrained by the downlink data rates and the satellite's onboard buffer storage. When the satellite is not in communication with a ground station, the remote sensing data is temporarily stored in a buffer until they can be offloaded. However, for many EO satellites, the amount of data generated often exceeds the buffer capacity, leading to data loss. To address this, satellite operators can increase the number of ground stations being utilized, though this is costly. Therefore, FSO technologies have been useful in enhancing the transmission speeds of remote sensing data, thereby alleviating data bottlenecks. Satellites in LEO experience short data transfer windows (under 15 minutes) and therefore have a smaller percentage data transferred (PDT). GEO data relays are favored in this regard because the LOS time is not limited by the satellite's orbit, enabling continuous, multipath links to the ground.

        The TerraSAR-X Earth observation (EO) satellite, launched in June 2007, carries a Tesat-built optical communication terminal (OCT) as a secondary payload under contract with the German Aerospace Center (DLR), achieving inter-satellite communication at 5.6 Gbps \cite{overton_2010, kampfner_2011}. In 2011, the Harbin Institute of Technology conducted China's first satellite-to-ground laser communication test, using the Ocean-2 satellite to achieve a downlink speed of 504 Mbps \cite{cae_2020}. The European Space Agency (ESA) developed the Small Optical Transponder (SOTA) project, which was mounted on NICT’s LEO Space Optical Communications Research Advanced Technology Satellite (SOCRATES) and launched in May 2014. This system demonstrated a space-to-ground link with a data transmission speed of 10 Mbps using a 50 kg-class small satellite \cite{kolev_2015}. Additionally, the EU Sentinel-1 (launched April 2014) and Sentinel-2 (June 2015) satellites employ onboard OCTs for communication with the European Data Relay System (EDRS). This system includes two payloads in GEO and utilizes an optical LEO-GEO link with a transmission speed of 2.8 Gbps (1.8 Gbps for user data) via Tesat's laser communication terminal (LCT) and a 300 Mbps Ka-band GEO-to-ground link \cite{esa_edrs_2015}. The EDRS includes five Sentinel satellites in LEO for data transmission \cite{esa_edrs_2024}.
        
        In 2019, JAXA launched the Japanese Data Relay System (JDRS), incorporating optical data relay satellites and JAXA’s Laser Utilizing Communication System (LUCAS) OCT for LEO-GEO communications, achieving speeds up to 1.8 Gbps \cite{satoh_2017}. Another significant step is ESA's HydRON project, aiming to establish the world’s first all-optical, multi-orbit transport network with terabit-per-second capacity, with the first satellite launch scheduled for 2026 \cite{jewett_2024}. In June 2023, China's Changguang Satellite Technology (CGST) and the Aerospace Information Research Institute (AIR) of the Chinese Academy of Sciences (CAS) achieved a 10 Gbps transmission speed test using the Jilin-1 MF02A04 EO satellite \cite{jones_china_2023}. The Jilin-1 constellation is expected to expand to 300 satellites by 2025, including models equipped with space-to-ground OCT. Satellogic, a Uruguayan EO data provider, is considering the deployment of OCTs in its next EO constellation \cite{satellogic_2023}. In September 2023, Satellogic contracted Skyloom's OCT, enabling the use of the Sky-Compass-1 GEO data relay network, which is expected to be operational in 2025.
        
        Capella Space, another EO data provider, is deploying Mynaric laser terminals on its Acadia satellites to operate with the US-based SpaceLink constellation in medium Earth orbit (MEO), which will receive data from LEO satellites via laser links and transmit it to the ground via RF signals. Originally, SpaceLink planned to offer data-relay services by early 2024; however, financial challenges led to the cessation of operations \cite{datarelay_2022}. Spire is also developing OCT inter-satellite links to be used with data-relay stations, while Warpspace is working on "WarpHub InterSat," the world’s first commercial optical inter-satellite communication service in MEO. Warpspace estimates that \$180 billion in EO data is lost due to connectivity gaps and plans to operate a LEO data-relay service by 2025, a cis-lunar network by 2030, and a Martian network by 2035 \cite{datarelay_2022}.
        
        Hedron, after raising \$17.8 million in Series A funding, intends to deploy the first orbital plane of what it describes as the world’s first hybrid optical/RF data relay network \cite{hedron_2024}. Astrophysics applications are also advancing, with NASA demonstrating a 200 Gbps space-to-ground optical link in April 2023 using the TBIRD system aboard NASA's PTD-3 satellite, marking a significant milestone in communication capabilities for future missions \cite{nasa_lcrd_overview, mit_lcrd_overview}.
        
        Currently, there is limited commercial use of space-to-ground FSO technology from EO satellites, with most systems relying on LEO-GEO inter-satellite link data relay systems, such as EDRS. However, even the EDRS employs a Ka-band downlink. Optical downlinks are still under development, with emerging technologies like Skyloom's Sky-Compass, ESA's HydRON network, WARPSPACE, and the Jilin-1 constellation expected to play key roles in future EO infrastructures.
                
    \subsection{Satellite Communications}

        Unlike EO systems, satellite communications require a bidirectional point-to-point data transmission link, where two ground stations transmit data in the form of voice and video telephone services, internet traffic, television broadcasting, command and control, and PNT through the same satellite system.
        
        Legacy GEO-located telecommunications satellites typically adopt numerous RF spot beams to provide TV broadcasting and direct-to-home broadband across large regions. However, Greece-based Hellas Sat is developing its GEO-located 'Hellas Sat 5' telecommunications satellite connected with feeder links in Greece, France, the ESA OGS network, and Thales Alenia Space's LEO HydRON satellites. Thales Alenia Space will develop the optical communication payload with very high data rate capacities (up to 1 Tbps) through the Vertigo H2020 programme.
        
        Satellite mega-constellations have benefited significantly from the adoption of OISL mesh networks enabling low-latency internet services \cite{tiwari_2020}. Notable examples include Starlink, Kuiper, OneWeb, Telesat Lightspeed, IRIS2, and Qianfan. However, the downlink transmission to ground stations in these systems typically uses RF (such as Ku-band) rather than optical. Among recent activities, the following are highlighted for the development of FSO in mega-constellations. In the United States, SpaceX’s Starlink satellites have been operating since the 2020s, and each satellite has four OISL terminals and Ku-band uplink/downlink antennas \cite{starlink_2021}. Amazon is developing its Project Kuiper broadband mega-constellation with OISL, with the first launch expected in late 2024 \cite{amazon_2023}. In Europe, Telesat is developing its 'Lightspeed' LEO constellation with Thales Alenia Space-manufactured OISL terminals \cite{thales_oisl_2021}. OneWeb’s second-generation satellites are yet to be launched but are said to include OISL \cite{oneweb}. The EU's 'IRIS2' satellite constellation multi-orbit broadband communications network will utilize the EuroQCI infrastructure and is led by satellite operators Eutelsat, SES, and Hispasat, along with manufacturers Airbus Defence and Space and Thales Alenia Space \cite{eu_iris2_2023, spacenews_iris2_2023}. Canadian-based Kepler Communications demonstrated OISL between two pathfinder satellites in LEO in June 2024 and is planning to adopt OISL on its space-to-space data-relay network called Kepler Network, which is compatible with the SDA's constellation \cite{breakingdefence_2024}.
        
        Deep-space communications have harnessed FSO communications for space exploration and scientific research. Given the high-bandwidth data transmission characteristic of optical communication, real-time or near-real-time data collection applications can optimize the decision-making process and data processing to improve the productivity of scientists on the ground \cite{trichili_2020}. NASA's Lunar Laser Communication Demonstration (LLCD) previously set a milestone by achieving a 622 Mbps downlink space-to-ground link from cis-lunar orbit, which in 2013 was the farthest space-to-ground FSO link ever achieved \cite{llcdmission_nasa}. An Artemis 2 Optical Communications System (O2O) is being integrated into the Orion spacecraft in preparation for the next Artemis 2 mission, which is expected to launch by the end of 2025, assuming no further programme delays \cite{artemis_laser_2023}. NASA's Psyche mission included the Deep Space Optical Communications (DSOC) experiment, which represents the first demonstration of deep-space optical communications. Launched in October 2023, the mission successfully delivered its first images in November 2023 and stands as the only deep-space optical mission to date \cite{nasa_dsoc_2023}.
        
        In summary, Hellas Sat 5 remains the only publicly advertised GEO telecommunications satellite to incorporate a space-to-ground optical feeder link, although it would not be unexpected if other legacy GEO SatCom operators follow suit. OISL mesh networks are nearing technological maturity on mega-constellations, and operators are increasingly adopting space-to-space data relay. Next-generation satellites, such as Starlink V2, are incorporating higher throughput RF bands such as Ka-band and E-band for space-to-ground links to keep up with growing throughput demands \cite{starlink_2021}. In lunar and deep-space applications, FSO links are increasingly being used both for direct-to-ground and relay systems.
        
    \subsection{Quantum Key Distribution}
    
        Quantum Key Distribution (QKD) is a secure communication method for exchanging encryption keys exclusively between shared parties \cite{orsucci_2024}. In entanglement-based QKD, a classical communication channel is used to transmit the necessary information for establishing a shared secret key, while a quantum channel transmits quantum states that carry the encoded information. Entanglement is a phenomenon where two or more quantum particles become correlated such that the state of one particle cannot be described independently of the others \cite{orsucci_2024}. Transmitting quantum signals through terrestrial optical fibers, however, is hindered by exponential absorption, significantly limiting the distance over which single-photon-level quantum signals can be transmitted.
        
        The first quantum communication technology demonstration took place in 2015 on the National University of Singapore's Galassia CubeSat (~2 kg) and was followed by the launch of Galassia-2 in July 2023. In August 2016, China and Austria launched the Micius small satellite (635 kg) to demonstrate space-to-ground QKD, achieving a downlink speed of 5.12 Gb/s between two quantum ground stations located 2,600 km apart \cite{zhang_2017, chen_2021}. SpeQtral and RAL Space are planning to launch the 12U SPEQTRE satellite in 2025 to demonstrate QKD from space using an 8 cm diameter aperture telescope. A follow-on mission, SpeQtral-1, is planned for launch later in the same year \cite{speqtral_2023}. In support of this, Archangel Lightworks and SpeQtral are collaborating to develop Quantum Lasercomms Optical Ground Stations (QLOGS) as part of the QLOGS Project \cite{archangel_speqtral_2023}. Scottish-based Craft Prospect secured grant funding from UKSA and ESA in 2020 to develop a QKD payload on a 6U CubeSat for the QKD Responsive Operations for Key Services (ROKS) pathfinder mission, with assistance from Bristol and Strathclyde Universities \cite{craft_prospect_2023}. The Qube Project, which demonstrated a 3U QKD satellite developed by a DLR-led consortium in Germany, was launched in August 2024. A follow-on mission, QUBE-II, is a 6U QKD CubeSat expected to launch in 2025 with a larger terminal aperture. The Jinan-1 satellite, smaller than Micius, was launched by China in 2022 and may be utilized in a quantum satellite constellation that integrates satellite QKD with terrestrial networks. In Europe, an SES-led consortium including TESAT, TNO, Airbus Netherlands, DEMcon, Celestia STS, and Officina Stellare is developing a European space-based QKD system (space and ground segments) as part of the European Quantum Communications Infrastructure (EuroQCI) \cite{ses_eagle1_2023}. The project is jointly funded by the Netherlands Space Office (NSO) and NXTGEN HIGHTECH, supporting future EuroQCI missions such as SAGA, IRIS2, TeQuantS, and Caramuel. An additional mission slated for late 2025 is the Platform for Optical Quantum Communications developed by the UK Quantum Communications Hub consortium (comprising York, Bristol, Heriot-Watt, and Strathclyde Universities, and RAL Space), which will demonstrate both discrete variable and continuous variable QKD. The Canadian QEYSSat mission, planned for 2025-26, aims to demonstrate an uplink scenario where the establishment of a shared key following the QKD protocol is done through the transmission of individual photons by a laser link from a ground station to a micro-satellite \cite{QEYSSat_web}.
        
        Emphasis on existing missions has included CubeSats as a way to demonstrate the technology. Larger projects are now underway with China's QLD constellation and the EAGLE-1 ESA mission. The QKD industrial ecosystem is rapidly evolving, with increased collaboration between academic institutions, government agencies, and private companies driving innovation and deployment. Key challenges such as enhancing transmission distances, improving key generation rates, and integrating QKD with existing communication infrastructure are being actively addressed. The commercialization of QKD is expected to grow, particularly in sectors requiring high-security communications, such as finance, defense, and critical infrastructure. Furthermore, advancements in satellite technology and miniaturization are facilitating the deployment of QKD capabilities on smaller satellites, expanding the potential for global secure communication networks.

    \subsection{Existing Optical Ground Station Networks}
    \label{sec: ground}
        
        Several major players have emerged in recent years in the ground segment. By breaking this down into large OGS and small OGS systems, different key players can be distinguished, as well as the typical characteristics of OGS. More explanation on this is included in the following subsections. The focus is on complete systems rather than sub-system developers, though key players such as sub-system manufacturers, ground station service providers, and major contract awardees are also considered. Many of these systems are designed with scalability to accommodate specific applications and mission requirements.
    
        ESA's European Optical Nucleus Network (EONN) includes OGS in Tenerife, Spain, operated by ESA; Almeria, Spain, operated by DLR; and Nemea, Greece, operated by KSAT \cite{ksat_2024}. Additionally, the Hague OGS in the Netherlands is operated by TNO \cite{aacclyde_tno_2024}.
        
        ESA's Network of Optical Stations for Data Transfer to Earth from Space (NODES) has an OGS (OGS-1) in Western Australia, which was installed in 2023, and a second OGS (OGS-2) being installed in Santiago, Chile. The Swedish Space Corporation was funded by the Swedish National Space Agency under the ARTES Scylight programme for the NODES project, and the station was built by Safran as part of its 'IRIS OGS' collection \cite{ssc_2023}.
        
        NASA’s JPL operates two OGS, with OGS-1 in Table Mountain, California, and OGS-2 in Hawaii, managed by the Space Communications and Navigation (SCaN) programme at NASA and JPL respectively, which supported the LCRD experiment \cite{mit_lcrd_overview}. NASA also operates a 0.7 m 'Low Cost Optical Terminal' at NASA’s Goddard Space Flight Center in Greenbelt, Maryland, which was installed in 2021.
        
        The DLR Institute of Communications and Navigation in Oberpfaffenhofen, Germany, has a 0.8 m OGS that became operational in 2021. DLR also operates an OGS in Truaen, Germany \cite{rattenbury_2024}. DLR has also developed the Small Optical Ground Stations Focal-Optics Assembly (SOFA), a small, fully integrated, low-cost instrument that is to be mounted on COTS or even existing astronomical telescopes, enabling optical communication capabilities. These are considered in the NZ network and used in the Australian free-space optical communications network. The SPACE Research Center of the University of the Bundeswehr Munich deployed a 0.6m Transportable Optical Ground Station by DLR in 2011 to enable temporary OGS operations.
        
        The ASA OGS in Graz, Austria, is a multi-functional laser and quantum communication facility, as well as a space debris observation and laser ranging facility for artificial space objects \cite{rattenbury_2024}. It was involved in QKD experiments with Micius.
        
        The National Institute of Information and Communications Technology (NICT) has four telescopes in Japan, including a 1.5m and 1m OGS in Kohanei, Tokyo; a 1m OGS in Kashima, Ibaraki; and a 1m OGS in Onna, Okinawa \cite{toyoshima_2006}.
        
        The Australian Space Agency, in collaboration with the Western Australian Government and the University of Western Australia, is developing the Australian free-space optical communications network, set to be operational by 2026 \cite{icrar_teranet_2024}. Locations include a 0.7m fixed-ground station at the University of Western Australia (TeraNet-1), Mingenew (TeraNet-2), and a 0.4m mobile station at New Norcia (TeraNet-3).
        
        A 0.7m Quantum OGS at Mount Stromlo Observatory in Canberra, Australia, is operated by the Australian National University \cite{rattenbury_2024}. Additional current sites include Perth and Adelaide. Potential sites include Dongara, Alice Springs, Mt Kent, Warkworth (NZ), Waiheke (NZ), Ardmore (NZ), Kopuwai (NZ), Lauder (NZ), Mt John (NZ), Birdlings Flat (NZ), Black Birch (NZ), and Urenui (NZ).
        
        China has deployed a 0.5 m OGS in the Pamir Plateau, Xinjiang, developed by the National Space Science Center in September 2023. The 1 m Xinglong Observatory of the Chinese Academy of Sciences played a key role in the Micius experiment. Other optical telescopes are located in Delinha, Qinghai; Nanshan, Xinjiang; Lijiang, Yunnan; and Ali, Tibet.

            \begin{figure}[t] % use [t] or [htbp]; avoid [H] in REVTeX
            \centering
            \includegraphics[width=\linewidth]{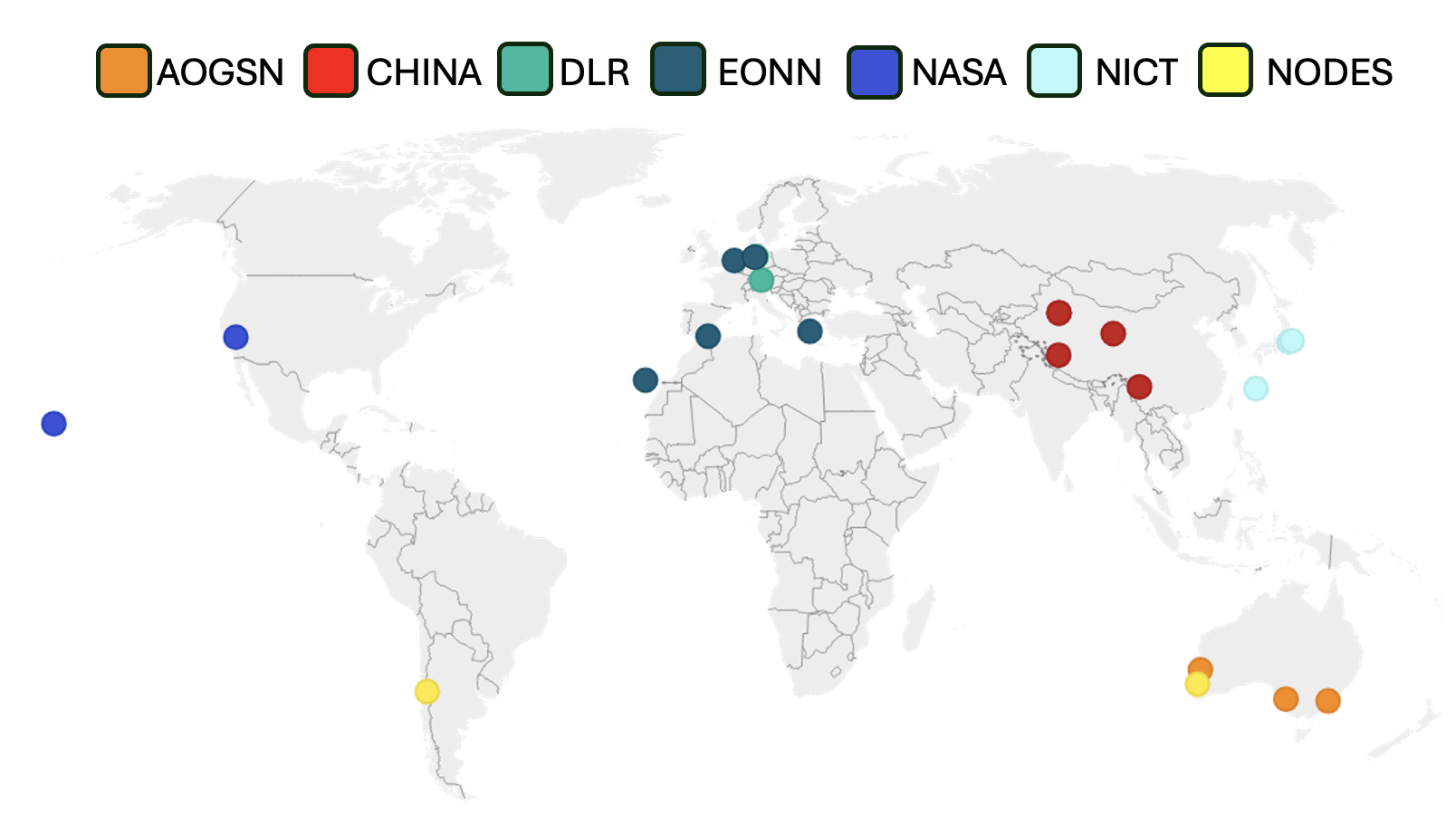}
            \caption{\textbf{Global distribution of key OGS networks.} This map highlights the locations of major OGS systems across various international initiatives, including ESA’s EONN, NASA’s SCaN network, China’s OGS network, and the AOGSN.}
            \label{fig:ogs_network}
            \end{figure}

    \subsection{Commercialisation in Optical Ground Stations}
    \label{sec: commerical}

        In the small OGS market, several key players have emerged from Europe. UK-based Archangel Lightworks is developing a small OGS named TERRA-M, as shown in Figure \ref{fig:smallogs} \cite{archangel_2024}. Archangel has collaborated with Japan’s Infostellar and SpeQtral to develop a mobile Quantum Optical Ground Station (Q-OGS) \cite{archangel_infostellar_2023, speqtral_2024}. Mynaric, a German startup, is working on FSO communication components for both space and ground segments, including an OGS that was recently awarded a contract by the Space Development Agency (SDA) \cite{myneric_2024}. Lithuania-based Astrolight offers its OGS-2 terminal, which can achieve a 12.5 Gbps downlink from a LEO satellite \cite{astrolight_web}. Cailabs, a French company, is developing the 0.8m TILBA-OGS, designed to deliver over 10 Gbps downlink from LEO satellites. Cailabs currently operates one TILBA-OGS in Rennes, France \cite{cailabs_2023}.
        
        In the United States, QinetiQ and General Atomics Synopta both advertise their OGS systems. The Aerospace Corporation operates three small OGS in California, Maui, and Albuquerque, which support both existing and future small satellite missions, aiming for an 800 Mbps downlink from CubeSats \cite{qinetiq_2020, ga-synopta, aerocorp_2022}.
        
\begin{figure}[H]
            
            \includegraphics[width=\linewidth]{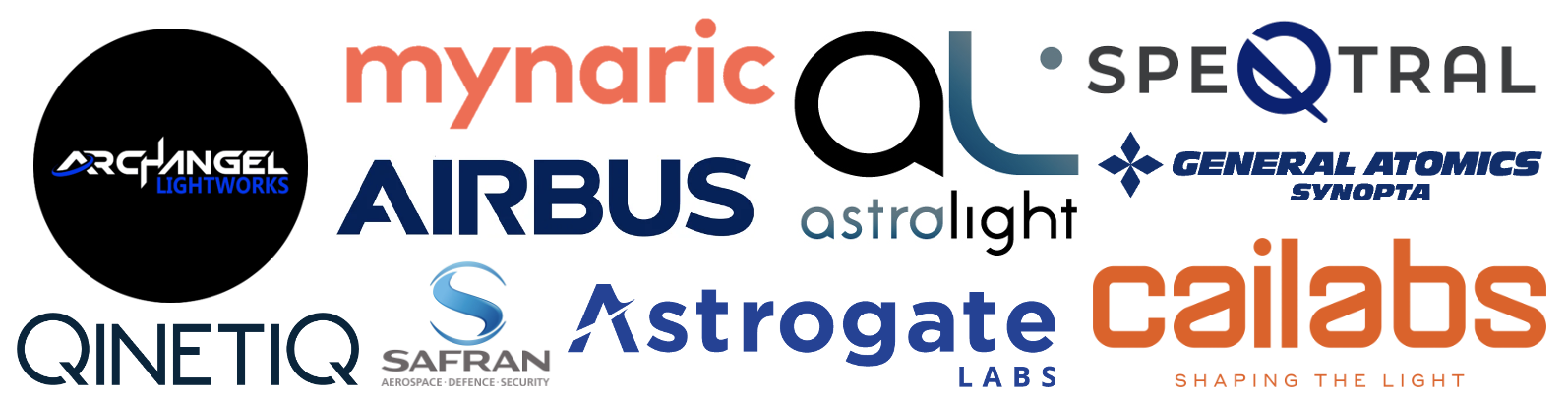}
            \caption{\textbf{Key commercial players in the OGS market.} This figure highlights major companies involved in the development and deployment of OGS systems, including leaders such as Airbus, Mynaric, General Atomics Synopta, and QinetiQ. These companies provide a variety of OGS solutions, from large-scale bi-directional links to small, portable stations used for satellite communication and quantum communication applications.
            }
            \label{fig:ogs-players}
        \end{figure}

        Small and mobile OGS with apertures under 1 m are dominating the commercial OGS market. These units, which cost several thousand dollars, can be trailer-mounted or even packed into suitcases, making them an attractive option for suppliers and adaptable within supply chain dynamics \cite{pettersson_2019}. Several of the small OGS systems surveyed for this paper also lack uplink capability entirely. Feeder links, however, are highly scalable, with uplink speeds exceeding 20 Gbps (scalable to Tbps) and downlink speeds surpassing 10 Gbps (also scalable to Tbps). Downlink rates vary significantly between suppliers, ranging from 1.25 Gbps for small OGS systems to 12.5 Gbps for larger ones. Tracking accuracy for OGS systems typically falls below 10 arcseconds. As illustrated in Figure \ref{fig:ogs-players}, several organizations have emerged as leaders in commercial OGS manufacturing and deployment services.

        Commercially available or in-development OGS include Airbus Netherlands, which offers the LaserPort OGS 100 Tb+ Class for 100 Gbps links with GEO. The LaserPort OGS 100 is a stationary bi-directional link to LEO satellites. General Atomics Synopta offers stationary and transportable OGS \cite{ga-synopta}. Contec offers OGS on the West Coast of Australia, built by Safran as part of its IRIS collection.

        \setlength{\parskip}{0pt}

        Large OGS systems typically feature aperture sizes greater than 1 m. The development and operation of larger OGS involve significant initial investments, particularly for larger apertures. These investments encompass terminal equipment, site facilities, wide-area communication infrastructure, weather and atmospheric monitoring systems, and aviation safety systems. In Europe, the cost estimation for an OGS is approximately \$2 million per station, with a range of \$1–5 million for LEO-to-ground link stations. This estimate includes maintenance and operating costs, which average around \$800,000 per station. The use of Commercial Off-The-Shelf (COTS) telescopes and mounts, along with custom facilities and fiber optics, is common practice to manage costs effectively \cite{schulz_2012}. Tracking accuracy for OGS systems generally falls below 10 arcseconds. The role commercialization plays is to mitigate these large costs, leading to the emergence of Ground Station as a Service (GSaaS), remote tasking, IoT-enabled minimization of human intervention and automation, and digital OGS solutions.

    \subsection{Global Overview and Conclusions}
        
        Within the FSO market, North America (the United States, Canada, and Mexico) is expected to hold a dominant share from 2022 to 2027 \cite{globalmarket_2024}. In North America, FSO is widely utilized in the aerospace and defense industries.
        
        The presence of existing research facilities in Western Europe and Eastern Europe is likely to support the growth of the free-space industry in these regions. Moreover, the Asia-Pacific region is projected to be the fastest-growing segment in the FSO market during the forecast period. This growth is driven by the region's increasing population, rising acceptance of sophisticated technologies and IoT, enhanced R\&D expenditure, and early adoption of innovative technologies. Furthermore, ongoing connectivity improvements and infrastructure renovations, along with increasing government investments, are fueling market expansion.
        
        Additionally, the integration of FSO technologies with emerging 5G networks and the expansion of smart city initiatives are expected to further boost demand in the Asia-Pacific region\cite {giiresearch_2024}\cite{transparencymarketresearch_2024}. The region's emphasis on digital transformation and sustainable development provides a conducive environment for the adoption of high-speed, secure FSO communication solutions. Furthermore, strategic partnerships between governments and private sector companies in research and development are likely to accelerate technological advancements and the deployment of FSO systems, enhancing the overall competitiveness of the global FSO market.
        
        Thus, the FSO market is poised for significant growth across multiple regions, driven by technological advancements, strategic investments, and the increasing demand for high-bandwidth, secure communication solutions in various industries.

\section{Network Simulation Methodology}
\label{chap: methodss}

\begin{figure}[H]
        \centering
        \includegraphics[width=\linewidth]{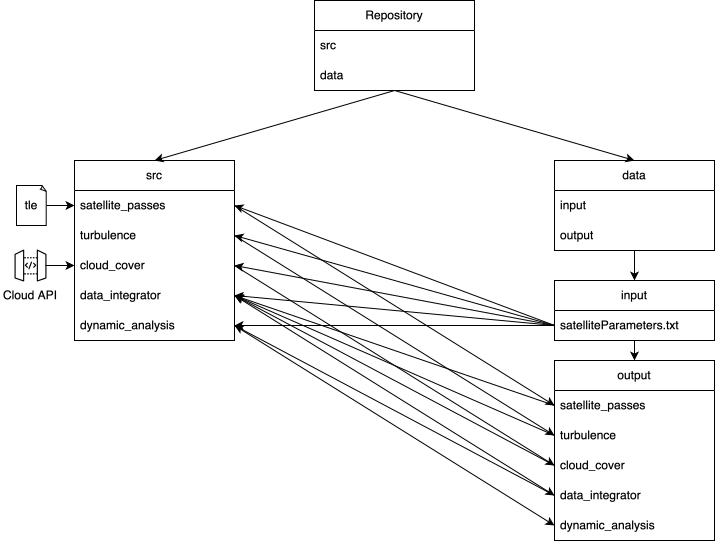}
        \caption{\textbf{System architecture for satellite pass prediction and data integration.} This diagram illustrates data flow among components like satellite pass predictions, turbulence, and cloud cover from TLE data and a Cloud API. The src directory processes and integrates inputs into the dynamic analysis pipeline, addressing network availability and data throughput.}
        \label{fig:software_architecture}
        \end{figure}

        The network simulation architecture is organized into two primary directories: \texttt{src} and \texttt{data}, as depicted in Figure~\ref{fig:software_architecture}. Within the \texttt{data} directory, the \texttt{input} subfolder contains the \texttt{satelliteParameters.txt} file, which serves as a fundamental input for the simulation scripts in the \texttt{src} subfolders. The \texttt{output} directory is divided into specific folders—\texttt{satellite\_passes}, \texttt{turbulence}, \texttt{cloud\_cover}, and \texttt{dynamic\_analysis}—each storing the results of their respective model components.
            
        The simulation workflow proceeds sequentially, starting with the \texttt{satellite\_passes} script, followed by \texttt{turbulence}, \texttt{cloud\_cover}, \texttt{data\_integrator}, and concluding with \texttt{dynamic\_analysis}. Each script within the \texttt{src} subfolders accesses necessary data from \texttt{satelliteParameters.txt}. The \texttt{data\_integrator} script aggregates outputs from the \texttt{satellite\_passes}, \texttt{turbulence}, and \texttt{cloud\_cover} modules. Subsequently, the \texttt{dynamic\_analysis} script utilizes both the \texttt{satelliteParameters.txt} and the integrated data from \texttt{data\_integrator} to perform comprehensive analyses.

        This structured approach facilitates the determination of Line of Sight (LOS) for Non-Geostationary Orbit (NGSO) satellites relative to a specified Optical Ground Station (OGS). Historical meteorological data are employed to assess the cloud-free LOS and estimate link degradation. The OGS performance is modeled through a link budget analysis, enabling the calculation of network availability, data throughput percentages, and data downlink capabilities. Network performance is then evaluated for Earth Observation (EO) applications.
        
        Subsequent sections provide a detailed breakdown of each component of the model, elaborating on the functionalities and methodologies employed within the \texttt{satellite\_passes}, \texttt{turbulence}, \texttt{cloud\_cover}, \texttt{data\_integrator}, and \texttt{dynamic\_analysis} modules, as well as the configurations of the OGS used to execute the simulations and obtain results.

    \subsection{Satellite Trajectory Modeling}
        
\begin{figure}[H]
            \centering
            \includegraphics[width=\linewidth]{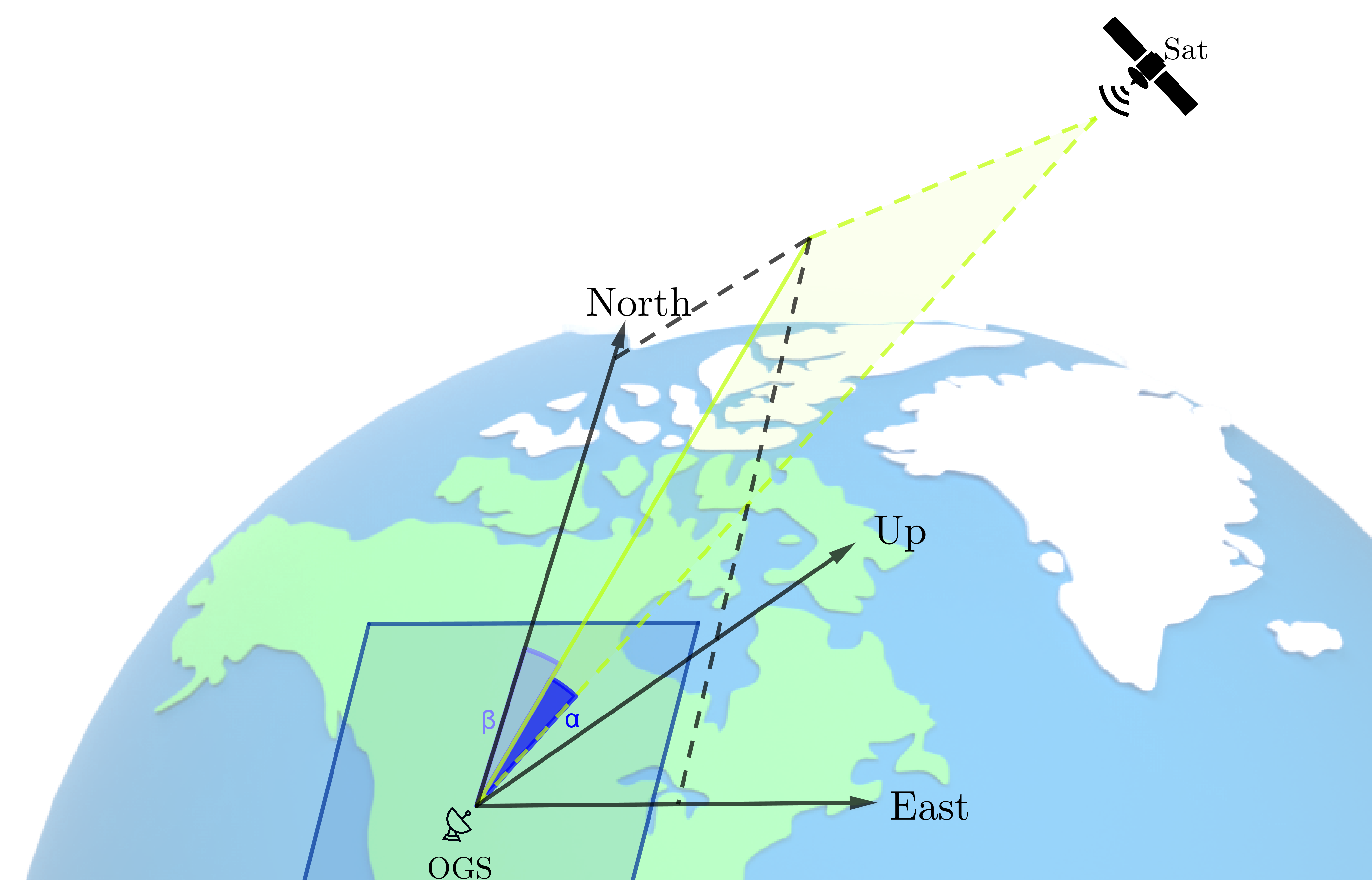}
            \caption{\textbf{Satellite trajectory modeling using TLE and Skyfield orbit propagator.} The satellite's trajectory is modeled using Two Line Element (TLE) files, and Skyfield estimates its position over time relative to the ground station.}
            \label{fig:geo_los}
            \end{figure}

        The trajectory of the satellite is modeled using orbital parameters defined in Two Line Element (TLE) files. The Skyfield orbit propagator is utilized to estimate and propagate the satellite's position over time, characterized by Azimuth ($\alpha$) and Elevation ($\beta$) angles relative to the ground station, as illustrated in Figure \ref{fig:geo_los}.

        Elevation refers to the angle between the satellite and the horizontal plane of the ground station. In practical scenarios, LOS obstructions such as mountainous terrain, vegetation, or surrounding structures of the OGS impose constraints on the minimum elevation angle at which a satellite remains visible, typically ranging between $5^{\circ}$ and $20^{\circ}$ \cite{rattenbury_2024}. The duration of a satellite pass is determined based on the total LOS time and the actual LOS interval, defined by the minimum elevation angle threshold. This approach ensures accurate modeling of satellite visibility and pass duration.
    \setlength{\parskip}{0pt}

    \subsection{Cloud Cover and Turbulence Modeling}
    
        EUMETSAT, the European operational satellite agency, is essential for monitoring weather, climate, and environmental conditions from space. Utilizing data from Meteosat GEO satellites and a constellation of  LEO satellites, EUMETSAT produces comprehensive data products that offer global meteorological insights. Among these is the Cloud Mask data \cite{eumetsat_cloud_mask}, which provides a binary representation of cloud cover over Europe with a spatial resolution of 3 km × 3km and a temporal resolution of 15 minutes, assigning a value of 2 for cloud presence and 0 for clear conditions.
        
\begin{figure}[H]
            \includegraphics[width=\linewidth]{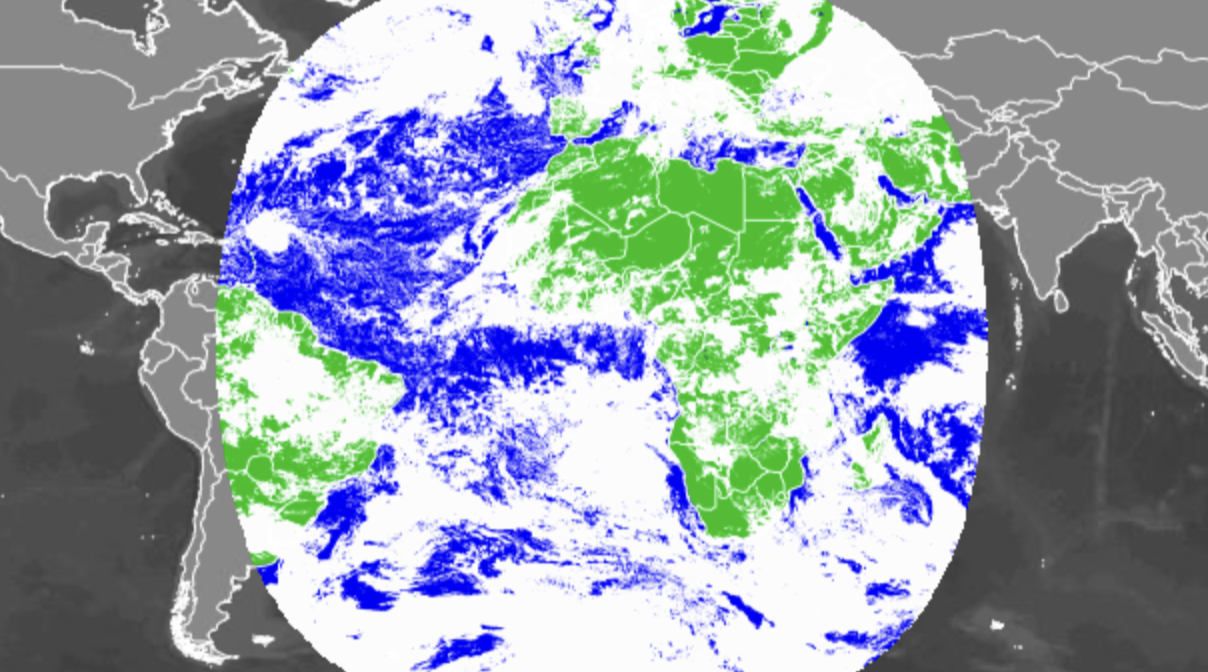}
            \caption{\textbf{Cloud cover data from EUMETSAT’s Meteosat GEO satellites.} The cloud mask data, limited to Africa, Europe, and parts of South America, indicating cloud cover (2) and clear conditions (0) \cite{eumetsat_cloud_mask}.}
            \label{fig:eumetsat}
            \end{figure}

        To quantify cloud cover over specific OGS, regions of interest were defined around each station using bounding boxes. The cloud cover metric was calculated by averaging the grid point values within these regions, utilizing OGS locations from the \texttt{satelliteParameters.txt} file. Cloud cover data from June 1, 2023, to June 1, 2024, were processed to assess the Cloud-Free Line of Sight (CFLOS) for each OGS. Network availability was determined based on the CFLOS, with outages classified when cloud cover exceeded 10\% across all OGS locations simultaneously. Figure \ref{fig:eumetsat} illustrates the spatial and temporal resolution of the EUMETSAT cloud cover measurements.

\begin{figure}[]
            \centering
            \includegraphics[width=\linewidth]{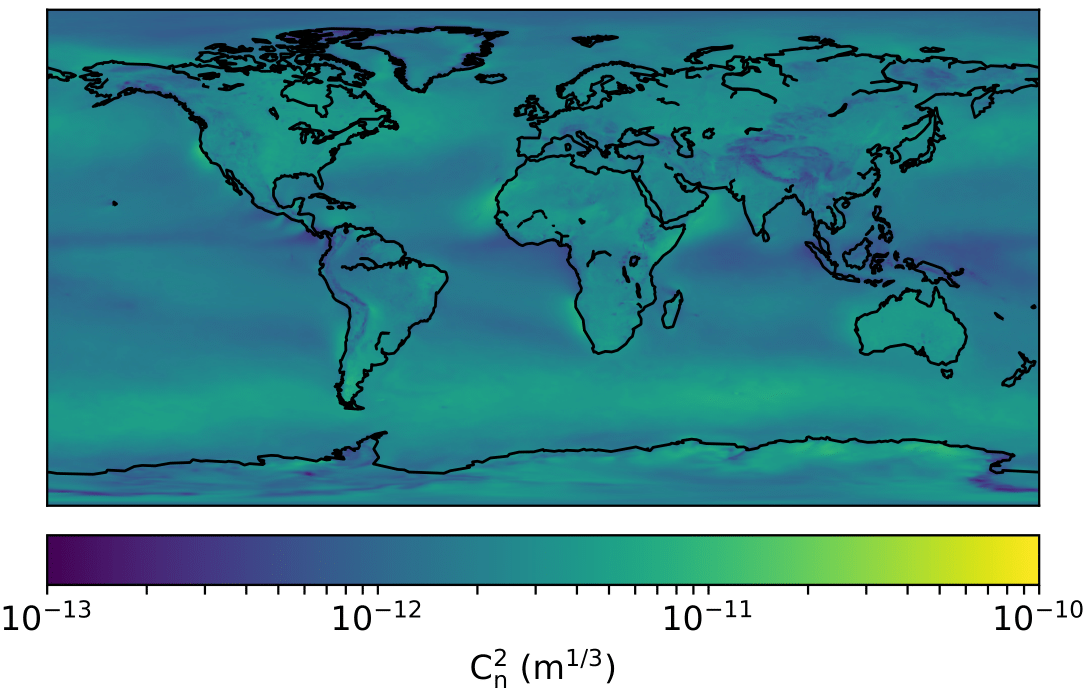}
            \caption{\textbf{Global map of atmospheric optical turbulence strength ($C_n^2$).} The scale represents turbulence strength in units of $m^{-1/3}$, where lower values (purple) indicate weaker turbulence, and higher values (yellow) represent stronger turbulence. \cite{osborne_2023}.}
            \label{fig:turbulence}
            \end{figure}

        Turbulence strength, a critical factor influencing the performance of optical communication systems, was modeled using annual average data provided by Dr. James Osborne. These C2 global maps of turbulence strength were derived from the European Centre for Medium Range Weather Forecasts (ECMWF) ERA-5 reanalysis dataset, offering a spatial resolution of approximately 10 km and temporal resolution of three-hour intervals for the year 2018 \cite{osborne_2023}. The turbulence modeling excluded contributions from local structures such as buildings or within the OGS dome and telescope, focusing solely on atmospheric turbulence. The turbulence data were integrated into the simulation by parsing turbulence strength values from FITS files corresponding to each OGS location. This integration enabled the evaluation of turbulence impacts on optical link performance, thereby informing site selection, instrument design, and the development of turbulence mitigation strategies. Figure \ref{fig:turbulence} presents an example of the turbulence values obtained from Dr. Osborne's dataset.

    \subsection{Data Throughput Analysis}
    
    \subsubsection{Methodology}
    
        The data throughput analysis aims to quantify the total volume of data transmitted from the satellite to ground stations over a specified period. This analysis accounts for several critical factors, including satellite passes, environmental conditions, and physical system parameters. Satellite passes refer to the instances when the satellite maintains a CFLOS with a ground station. Environmental factors encompass cloud cover, atmospheric turbulence, and atmospheric attenuation, all of which can significantly impact data transmission quality. Additionally, physical parameters such as antenna gains, free-space path loss, pointing losses, and system noise are integral to determining the achievable data rates during each satellite pass.
        
        By integrating these factors, the analysis calculates the achievable data rate for each satellite pass and subsequently aggregates these rates to determine both monthly and overall data throughput.
    
    \subsubsection{Formulas Employed}

\paragraph{Link Budget Equation}
The link budget equation is fundamental in calculating the received power (\( P_{\text{rx}} \)) at the ground station:
\begin{equation}
\label{eq:link_budget}
    P_{\text{rx}} = P_{\text{tx}} + G_{\text{tx}} + G_{\text{rx}} - L_{\text{path}} - L_{\text{other}}
\end{equation}
\cite{mathworks_link_budget}

where:
\begin{itemize}
    \item \( P_{\text{tx}} \) is the transmitted power (dBW).
    \item \( G_{\text{tx}} \) and \( G_{\text{rx}} \) are the transmitter and receiver antenna gains (dB), respectively.
    \item \( L_{\text{path}} \) represents the path loss (dB).
    \item \( L_{\text{other}} \) encompasses other losses (dB), including pointing loss, atmospheric attenuation, cloud attenuation, turbulence loss, and link margin.
\end{itemize}

\paragraph{Free-Space Path Loss}
The free-space path loss (\( L_{\text{fs}} \)) quantifies the loss of signal strength as it propagates through free space:
\begin{equation}
\label{eq:free_space_loss}
    L_{\text{fs}} = 20 \log_{10}\!\left( \frac{4\pi S}{\lambda} \right)
\end{equation}
\cite{fspl}

where:
\begin{itemize}
    \item \( S \) is the slant range distance (m) between the satellite and ground station.
    \item \( \lambda \) denotes the operational wavelength (m).
\end{itemize}

\paragraph{Antenna Gain}
Antenna gain (\( G \)) for both the satellite and ground station is calculated using:
\begin{equation}
\label{eq:antenna_gain}
    G = 10 \log_{10} \!\left( \eta \left( \frac{\pi D}{\lambda} \right)^2 \right)
\end{equation}
\cite{FCC2017}

where:
\begin{itemize}
    \item \( \eta \) represents antenna efficiency (dimensionless).
    \item \( D \) is the antenna aperture diameter (m).
\end{itemize}

\paragraph{Pointing Loss}
Pointing loss (\( L_{\text{p}} \)) arises from misalignment between the satellite and ground station antennas and is given by:
\begin{equation}
\label{eq:pointing_loss}
    L_{\text{p}} = -10 \log_{10} \!\left( e^{- \left( \frac{2 \sigma_{\theta}}{\theta_{\text{div}}} \right)^2 } \right)
\end{equation}
\cite{Pavlak2017}

where:
\begin{itemize}
    \item \( \sigma_{\theta} \) is the combined pointing error (rad).
    \item \( \theta_{\text{div}} \) is the beam divergence angle (rad).
\end{itemize}

\paragraph{Atmospheric Attenuation}
Atmospheric attenuation (\( L_{\text{atm}} \)) as a function of the elevation angle (\( \theta_{\text{elev}} \)) is modeled by:
\begin{equation}
\label{eq:atmospheric_attenuation}
    L_{\text{atm}} = \gamma \cdot \frac{1}{\sin(\theta_{\text{elev}})}
\end{equation}
\cite{ITU1997}

where \( \gamma \) represents the specific attenuation (dB/km).

\paragraph{Signal-to-Noise Ratio (SNR)}
The Signal-to-Noise Ratio (SNR) in decibels is calculated using:
\begin{equation}
\label{eq:SNR}
    \text{SNR}_{\text{dB}} = P_{\text{rx}} - \left( k_{\text{B,dB}} + T_{\text{s,dB}} + 10 \log_{10} B \right)
\end{equation}
\cite{Bakulin2022}

where:
\begin{itemize}
    \item \( k_{\text{B,dB}} \) is the Boltzmann constant in dB.
    \item \( T_{\text{s,dB}} \) denotes the system noise temperature in dB.
    \item \( B \) is the bandwidth (Hz).
\end{itemize}

\paragraph{Shannon–Hartley Theorem}
The Shannon–Hartley theorem provides the maximum achievable data rate (\( C \)):
\begin{equation}
\label{eq:shannon_hartley}
    C = B \log_{2}(1 + \text{SNR})
\end{equation}
\cite{Hossain2019}

with SNR expressed in linear scale.

    \subsubsection{Implementation}
    
        The data throughput analysis is implemented in Python, utilizing libraries such as \texttt{pandas}, \texttt{numpy}, and \texttt{matplotlib}. The workflow begins with the initialization of satellite and ground station parameters, which are imported from a configuration file. Subsequently, combined pass data, including timings, durations, maximum elevations, cloud cover, and turbulence levels, are ingested into the system.
        
        For each satellite pass, the slant range distance is calculated using geometric models, which is then used to compute the free-space path loss. Antenna gains for both the transmitter and receiver are determined based on the antenna dimensions. Pointing loss is assessed by evaluating the pointing errors against the beam divergence angle. Atmospheric and environmental losses, including atmospheric attenuation, cloud attenuation, and turbulence losses, are subsequently calculated. These loss components are aggregated to obtain the total loss, which is then used to determine the received power via the link budget equation.
        
        The SNR is computed in decibels and subsequently converted to a linear scale to apply the Shannon-Hartley theorem, thereby determining the maximum achievable data rate for each pass. The data transmitted during each pass is calculated by multiplying the achievable data rate by the duration of the pass.
        
        Aggregated data is then compiled to provide monthly and overall data throughput metrics. Visualization tools generate graphical representations of monthly data transmission and per-pass data volumes. Finally, the aggregated data and visualizations are exported to CSV files for further analysis.
        
    \subsubsection{Result Calculation}

The total data transmitted is determined by summing the data transmitted during each satellite pass:
\begin{equation}
\label{eq:total_data_transmitted}
    T = \sum_{i} \left( C_i \times t_i \right)
\end{equation}
\cite{Burgin2008}

where:
\begin{itemize}
    \item \( C_i \) is the achievable data rate during pass \( i \) (bps).
    \item \( t_i \) is the duration of pass \( i \) (s).
\end{itemize}

The maximum possible data transmitted under ideal conditions is calculated as:
\begin{equation}
\label{eq:maximum_data_transmitted}
    M = \sum_{i} \left( C_{\text{max}} \times t_i \right)
\end{equation}
\cite{rioul2014shannon}

where:
\begin{itemize}
    \item \( C_{\text{max}} \) is the maximum achievable data rate under ideal conditions (bps).
    \item \( t_i \) is the duration of pass \( i \) (s).
\end{itemize}

Here, \( C_{\text{max}} \) is directly related to the Shannon capacity, which defines the maximum data rate that can be achieved over a communication channel with a given bandwidth and SNR without error, assuming ideal conditions. This parameter serves as a benchmark to evaluate the efficiency and performance of the actual data transmission under realistic operational scenarios.

The percentage data throughput is then derived:
\begin{equation}
\label{eq:percentage_throughput}
    \frac{T}{M} \times 100\%
\end{equation}

where:
\begin{itemize}
    \item $T$ = Total Data Transmitted
    \item $M$ = Maximum Possible Data Transmitted
\end{itemize}

        All of these calculations collectively provide insights into per-pass data volumes, monthly aggregates, and overall performance metrics which are essential parameters to consider for real-world deployment. 
    
    \subsection{Network Availability Analysis}
    
    \subsubsection{Methodology}
    
        The network availability analysis assesses the proportion of time during which the satellite maintains a viable communications link with at least one ground station, taking into account cloud cover conditions. This analysis leverages real-time cloud cover measurements at each ground station and employs a predefined cloud cover threshold to determine the feasibility of communication. The temporal resolution of the analysis ensures that availability is evaluated at each discrete time instance within the dataset.
    
                \noindent The availability at a specific time (\(t\)) is defined as:
        \begin{equation}
        \label{eq:availability}
        A(t)=
        \begin{cases}
        1, & \text{if } \exists\,\text{GS}_i \text{ such that } \text{CC}_i(t) < \text{th},\\[2pt]
        0, & \text{otherwise}.
        \end{cases}
        \end{equation}
        
        \noindent where:
        \begin{itemize}
            \item \( \text{th} \) is the cloud-cover threshold,
            \item \( \text{CC}_i(t) \) is the cloud cover at ground station \( i \) at time \( t \).
        \end{itemize}
        
        \noindent The monthly network availability is calculated as:
        \begin{equation}
        \label{eq:monthly_availability}
        A_{\text{month}} = \frac{1}{N_{\text{month}}}\sum_{t \in \text{Month}} A(t)\times 100\%,
        \end{equation}
        \noindent where \(N_{\text{month}}\) is the total number of time instances in the month. \cite{Reliasoft2022}

    \subsubsection{Implementation}
    
        The network availability analysis is executed through a systematic process. Initially, cloud cover data for each ground station is imported and synchronized based on their timestamps to ensure temporal alignment. For each time instance, the analysis evaluates whether any ground station has cloud cover below the defined threshold. If at least one ground station satisfies this condition, an availability value of 1 (available) is assigned; otherwise, a value of 0 (unavailable) is recorded.
        
        Subsequent to availability determination, the data is grouped by calendar month to compute monthly availability metrics using the aforementioned formula. To maintain data integrity, incomplete months at the beginning and end of the analysis period are excluded from the aggregation. Visualization tools are then employed to create plots that illustrate monthly network availability trends, providing a clear depiction of the system's reliability over time. Finally, the monthly availability metrics and overall results are exported to CSV files for documentation and further analysis.
    
    \subsubsection{Result Calculation}

        \noindent Let \(A_i\) be the availability for month \(i\) and \(N\) be the total number of months. The overall network availability is:
        \begin{equation}
        \label{eq:overall_availability}
        A_{\text{overall}} = \frac{1}{N}\sum_{i=1}^{N} A_i
        \end{equation}
        \cite{Reliasoft2022}

        This metric offers a comprehensive overview of the satellite communication system's reliability throughout the analysis period, incorporating environmental conditions that may influence communication link viability.

    \subsection{Simulation Setup}
    
\begin{figure}[H]  \includegraphics[width=1\linewidth]{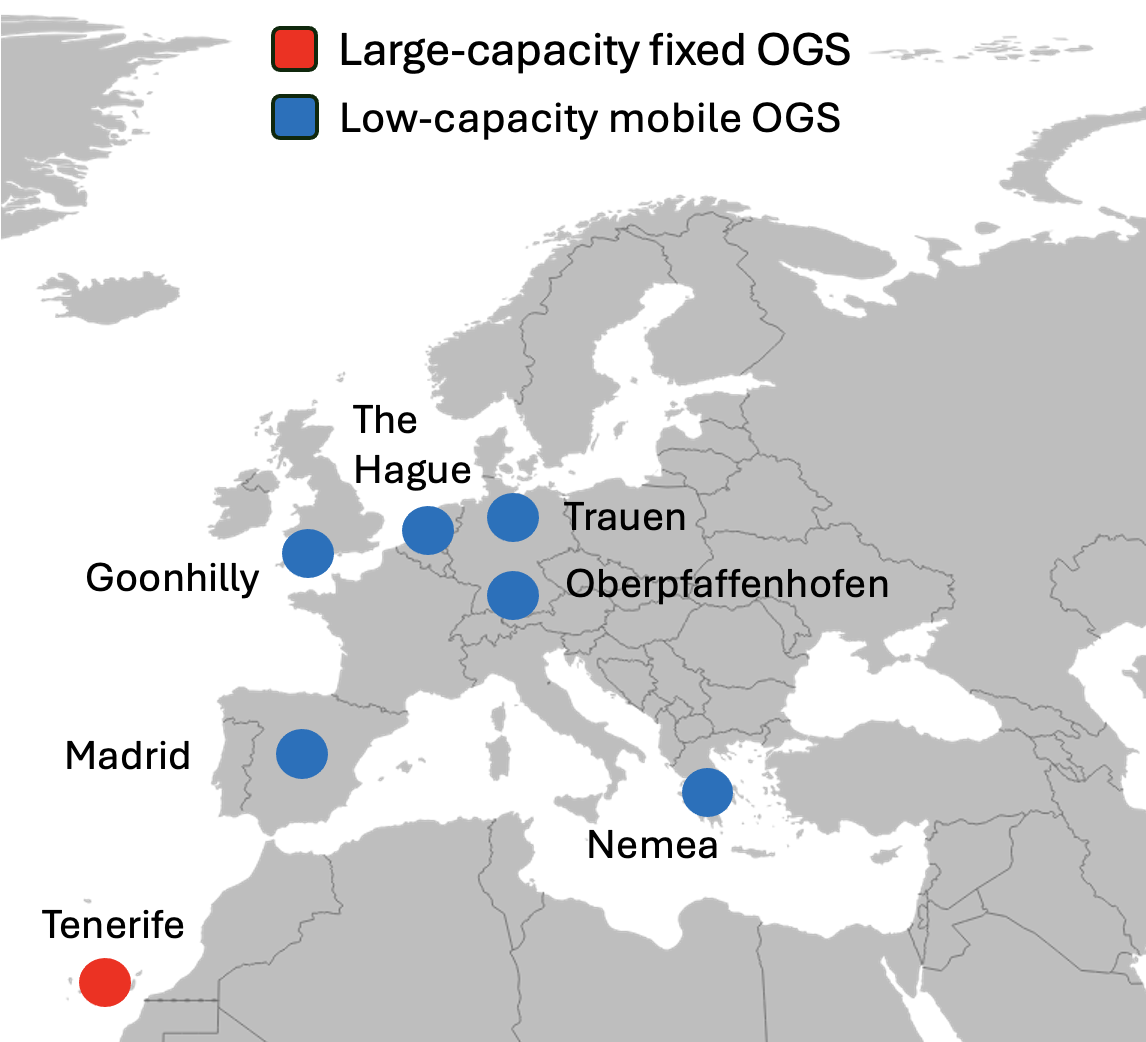} \caption{ \textbf{Pre-selected locations for OGS within the European Optical Nucleus Network (EONN).} [Red] markers represent larger capacity OGS, while [Blue] markers denote mobile terminals. The map highlights the geographical distribution across Europe, emphasizing strategic placement to optimize coverage and network resilience. Key sites are chosen based on factors such as elevation, climate stability, and proximity to existing infrastructure to minimize potential link outages and facilitate integration with terrestrial networks.} \label{fig:ogslocation} 
        \end{figure}

        Figure \ref{fig:ogslocation} illustrates the Optical Ground Stations (OGS) selected for this study. These locations are integral components of the European Optical Nucleus Network (EONN) and have been employed in numerous Free Space Optics (FSO) experiments, providing a robust foundation for our analysis. Additionally, these stations possess validated availability statistics, which are critical for benchmarking and verifying the overall accuracy and reliability of our simulation model
        
        A Line-of-Sight (LOS) analysis is conducted for a single satellite with an orbit identical to that of the TerraSAR-X satellite, which operates at a 514 km altitude with a $97.44^\circ$ inclination in a Sun-Synchronous Orbit (SSO) \cite{terrasarx_2009}. The TerraSAR-X payload generates 1.2 Tb/day of data, with an onboard storage capacity of 390 Gb and an RF downlink rate limited to 262 Mbps. In this experiment, the satellite is equipped with an Optical Communication Terminal (OCT) capable of transmitting data to the OGS at 1 Gbps.
        
        A configuration consisting of one large-capacity receiving OGS located in Tenerife, Spain, is compared with a network of six small receiving OGS located in Tenerife, Spain; Nemea, Greece; Trauen, Germany; Goonhilly, United Kingdom; Madrid, Spain; Oberpfaffenhofen, Germany; and The Hague, Netherlands.
        
        \subsection{Configurations Overview}
        
        To compare deployment options, we consider four configurations that combine a single large-capacity OGS in Tenerife with an increasing number of mobile stations (Table~\ref{tab:configs}). \textbf{Config 1} uses only Tenerife. \textbf{Config 2} augments Tenerife with two mobile stations: Trauen and Nemea. \textbf{Config 3} further adds Goonhilly and Madrid. \textbf{Config 4} includes Tenerife plus all six mobile stations—Trauen, Nemea, Goonhilly, Madrid, Oberpfaffenhofen, and The Hague. Tenerife is the large OGS; all others are mobile terminals.
        
        \begin{table}[H]
        \centering
        \setlength{\tabcolsep}{4pt}
        \renewcommand{\arraystretch}{1.05}
        \begin{tabular}{@{}l p{0.78\columnwidth}@{}}
        \hline
        \textbf{Config} & \textbf{Stations Included} \\
        \hline
        1 & Tenerife (Large) \\
        2 & Tenerife (Large); Trauen; Nemea \\
        3 & Tenerife (Large); Trauen; Nemea; Goonhilly; Madrid \\
        4 & Tenerife (Large); Trauen; Nemea; Goonhilly; Madrid; Oberpfaffenhofen; The Hague \\
        \hline
        \end{tabular}
        \caption{\textbf{Summary of Experimental OGS Configurations.} Tenerife is the large-capacity OGS; all others are mobile stations.}
        \label{tab:configs}
        \end{table}

\section{Results}
    
    Cloud cover is one of the primary limitations for optical communications from space to Earth. Atmospheric conditions can severely degrade or completely obstruct the optical signal, making it imperative to strategically select the locations of OGS. In our proposed network configurations, we have chosen a set of geographically dispersed sites to minimize the likelihood of simultaneous cloud coverage affecting all stations.  

    \begin{figure}[H]
    \includegraphics[width=\linewidth]{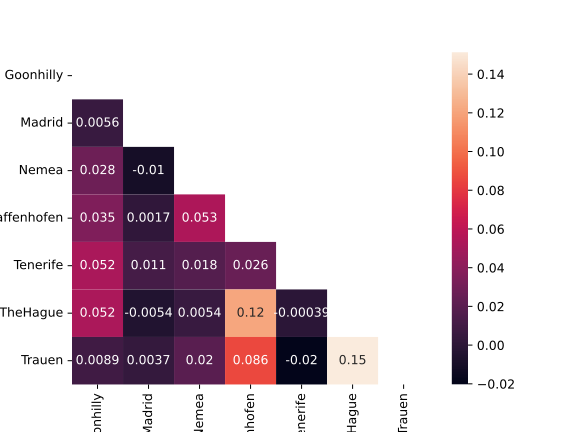}
    \caption{\textbf{Cloud-cover correlations across OGS sites.} Entries show the Pearson coefficient \(r\) between the cloud-cover time series for each site pair. Color encodes correlation strength (bar at right; \(-1\) to \(+1\)). Off-diagonal values are generally small (many \(r<0.02\); all \(|r|\lesssim 0.15\)), indicating weak synchrony of cloudy/clear conditions between locations. This supports a site-diversity strategy: when one station is weathered out, others are likely to remain clear, improving end-to-end link availability and reducing coincident outages.}
    \label{fig:pearson}
\end{figure}

    To assess the effectiveness of our site selection, we analyzed the Pearson correlation coefficients for cloud cover between each pair of OGSs over a year of data from June 2023 to June 2024. The Pearson correlation coefficient, $r$, quantifies the linear relationship between two variables—in this case, the cloud cover at two different OGS locations \cite{cloud_cover_pearson}. Low or negative correlation values indicate that the cloud cover patterns at the two sites are independent or inversely related, which is desirable for maintaining continuous optical communication links.
    
    As shown in Figure \ref{fig:pearson}, several site pairs exhibit low correlation values with $r < 0.02$, suggesting that cloud cover occurrences are largely uncorrelated between these locations. This implies that when one station is experiencing poor atmospheric conditions, the others are likely to remain operational, thereby enhancing the overall reliability of the network.

    In contrast, the sites in The Hague, Trauen, and Oberpfaffenhofen exhibit higher pairwise correlations, with positive values slightly above $0.1$. This increased cloud cover correlation is likely due to their geographical proximity and similar climatic conditions. Although still relatively low, it suggests a greater likelihood of simultaneous cloud cover events among these stations compared to more distant pairs.

    \begin{figure}[H]
    \includegraphics[width=\linewidth]{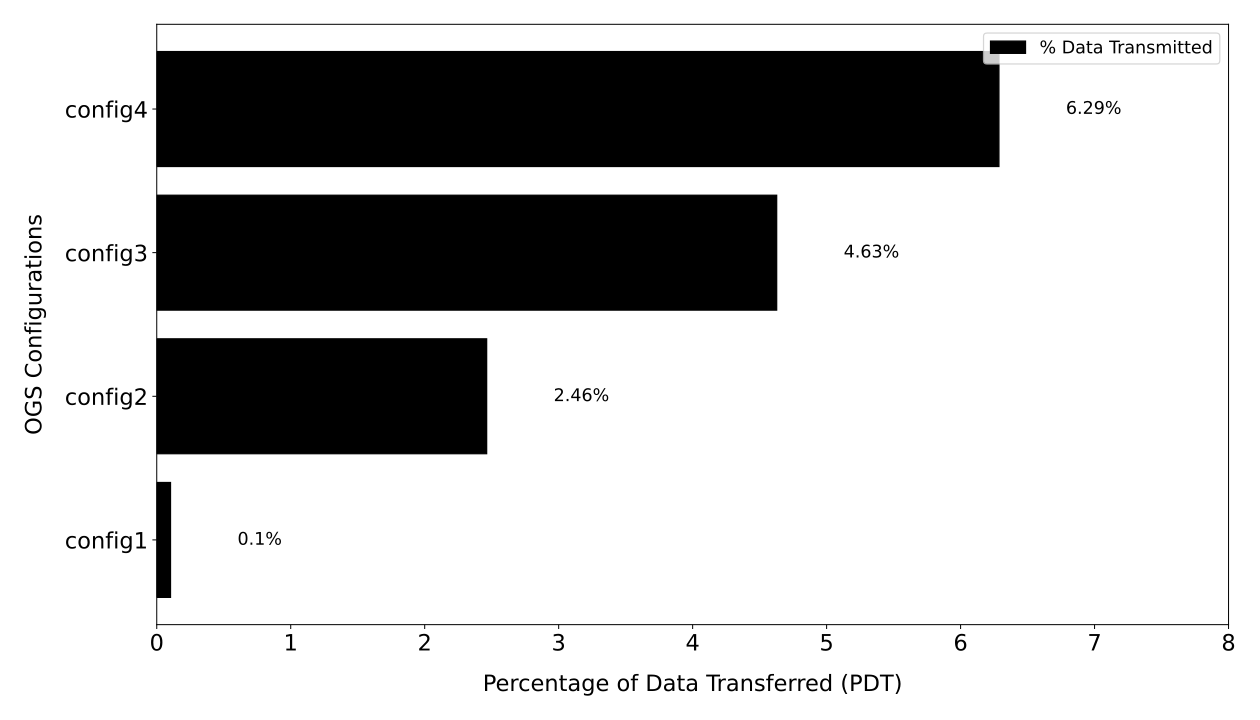}
    \caption{\textbf{Normalized Percentage of Data Transferred (PDT) across OGS network sizes.} 
    PDT is computed over the full campaign as the total downlinked payload divided by a fixed upper bound equal to the maximum attainable volume under uninterrupted visibility for the seven-site topology (config4). Normalizing to a common bound enables fair comparison of topologies independent of absolute link-budget assumptions.}
    \label{fig:PDT}
\end{figure}

    Building on these findings, Figure \ref{fig:PDT} shows the Percentage of Data Transferred (PDT) normalized to the maximum data transmitted for config4. The figure demonstrates that the network's data transfer capacity improves as more OGS are added. Notably, the network with only one ground station in Tenerife achieves a PDT of merely 0.1\%, highlighting significant data transfer limitations due to frequent outages.
    \setlength{\parskip}{0pt}

    In contrast, expanding the network to include additional ground stations results in notable improvements. For instance, the configuration with two ground stations raises the PDT to 2.46\%, while four ground stations further increase the PDT to 4.63\%. The most extensive configuration with six mobile ground stations reaches the highest PDT of 6.29\%. These results illustrate the benefits of geographic diversity; by spreading ground stations across different climatic regions, the network mitigates the risk of simultaneous weather-related outages. This increased availability ensures that even if one or more stations experience unfavorable conditions, others can compensate, maintaining continuous data flow.
    \setlength{\parskip}{0pt}

\begin{figure}[H]
    \includegraphics[width=\linewidth]{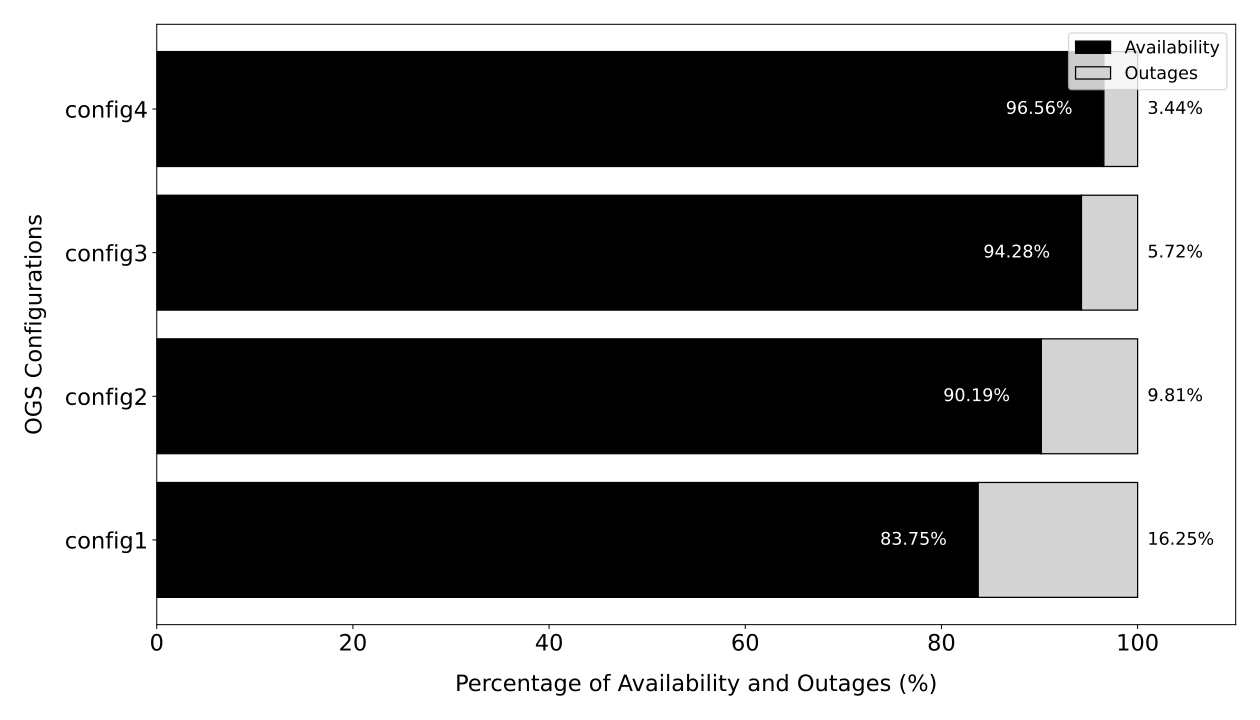}
    \caption{\textbf{Annual mean availability and outage frequency versus OGS network size.}
    Bars show the fraction of simulated time with at least one site able to support an optical downlink (availability, \(A\)) and its complement (outage, \(1-A\)), aggregated over a year-long campaign. We compute
    \(A=\tfrac{1}{T}\sum_{t}\mathbf{1}\{\exists\,\text{OGS meeting visibility and weather constraints}\}\).
    Availability increases with additional ground stations—reflecting reduced weather co-outages and more contact opportunities—but with diminishing returns. A single-station network (Tenerife) exhibits an outage rate of \(\approx 16.25\%\); adding two mobile sites reduces outages to \(\approx 9.81\%\), with further improvements for four and six sites.}
    \label{fig:outages}
\end{figure}

    The percentage of mean availability and outage over a year has also been evaluated, as shown in Figure \ref{fig:outages}. As expected, network availability improves with an increasing number of ground stations. For a single-station setup in Tenerife, the outage rate is around 16.25\%, indicating minimal availability. When the network expands to include two stations, the outage percentage drops to 9.81\%, showing moderate improvement.
    \setlength{\parskip}{0pt}

    Further expansion to three stations reduces the outage rate to 5.72\%, while the most extensive configuration with four ground stations brings the outage percentage down to 3.44\%. This progressive decline in outages demonstrates the advantages of deploying more OGSs across diverse geographical locations. By distributing ground stations across different climatic zones, the network mitigates the likelihood of simultaneous weather-induced outages, enhancing overall availability.
    \setlength{\parskip}{0pt}

\begin{figure}[H]
    \includegraphics[width=\linewidth]{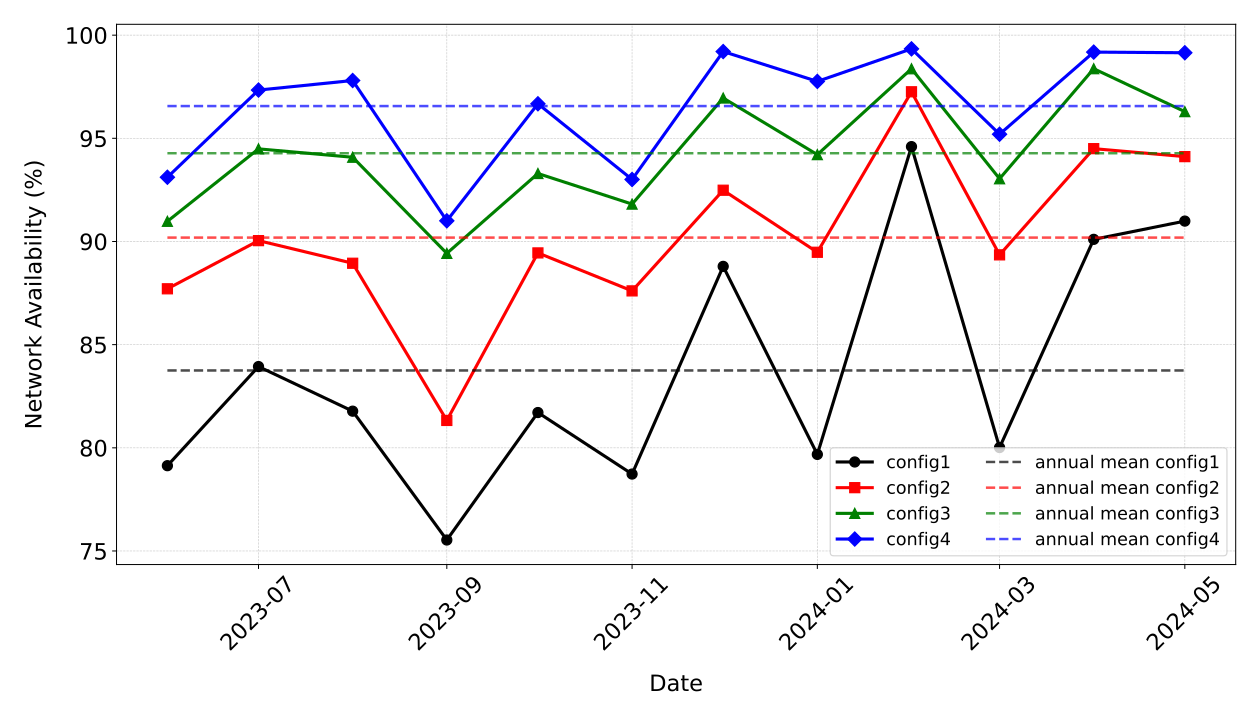}
    \caption{\textbf{Monthly mean network availability by OGS configuration.}
    Each curve reports, for a given topology, the fraction of simulation time in which at least one ground station meets the visibility constraint, averaged over each calendar month. The single-site network (config1) shows the lowest and most volatile availability—with recurring dips below \(\sim\!85\%\)—reflecting sensitivity to local cloud cover. Adding geographically separated sites raises the mean and reduces month-to-month variability, consistent with the weak inter-site cloud correlations in Fig.~\ref{fig:pearson}. Gains are sub-linear, indicating diminishing returns from additional stations beyond four to six sites.}
    \label{fig:network_availability2}
\end{figure}

    Figure \ref{fig:network_availability2} shows the monthly mean availability for each OGS network configuration. The graph reveals significant variability in availability for different configurations. The single ground station configuration (config1) displays the lowest availability, with frequent fluctuations dropping below 85\% during certain months, reflecting its high sensitivity to local weather conditions.

    As more ground stations are added, monthly availability profiles improve. Configurations 2 and 3 maintain availability mostly above 90\%, with slight monthly variations. Configuration 4, featuring the most stations, achieves the highest and most consistent availability, generally exceeding 95\% throughout the year. This highlights the network's enhanced resilience and reduced dependence on individual station conditions against localized weather disruptions.
    
\begin{figure}[]
    \includegraphics[width=\linewidth]{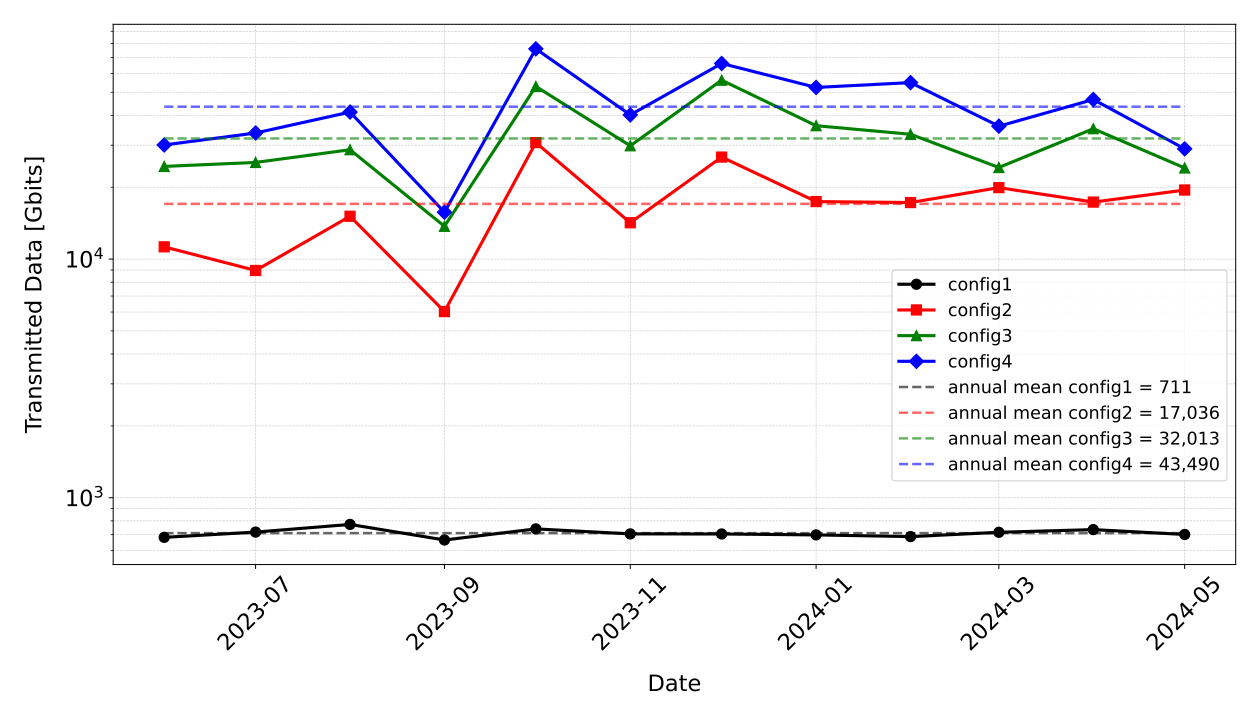}
    \caption{\textbf{Monthly transmitted data by network configuration (Gbits).}
    Totals are aggregated across sites; adding OGSs increases delivered payload and stabilizes month-to-month throughput—config1 remains below \(10^3\) Gbits, while configs 2–4 deliver progressively higher, steadier volumes (config4 highest).}
    \label{fig:tx_data}
\end{figure}

    Figure \ref{fig:tx_data} displays the monthly data transmitted for each network configuration. A significant increase is observed when transitioning from a single ground station (config1) to multiple stations. Config1 consistently transmits the least data, remaining below $10^3$ Gbits, which reflects its limited capacity and high reliance on the availability of a single station.
    
    In contrast, the configurations with additional ground stations (config2, config3, and config4) show significant improvements. Config2 and config3 exhibit steady increases in transmitted data, with annual means of 17,066 Gbits and 32,013 Gbits, respectively. Config4, with the most ground stations, achieves the highest data transmission rates, maintaining an annual mean of 43,490 Gbits and showing peaks that align with periods of improved network availability.

\begin{figure}[H]
        \includegraphics[width=\linewidth]{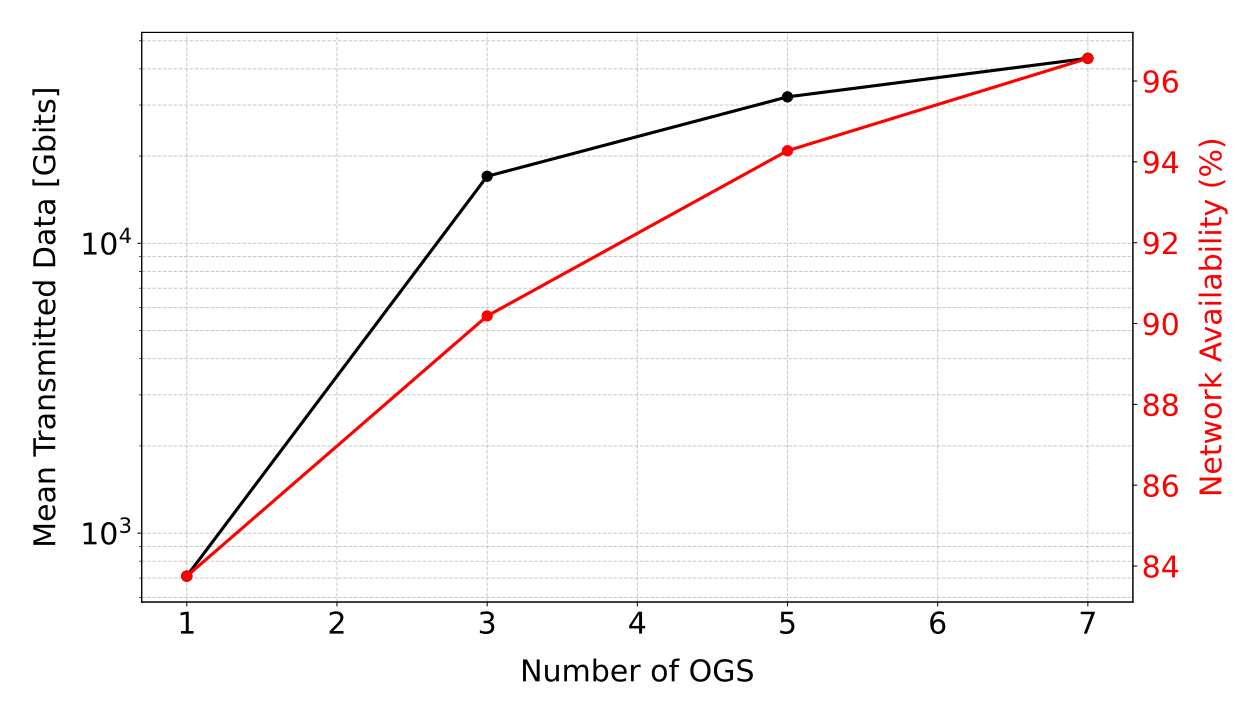}
        \caption{\textbf{Annual mean transmitted data vs.\ annual mean availability.}
        Each point corresponds to a network configuration (config1–config4). Both metrics rise as OGS count increases, yielding a clear positive correlation; gains are monotonic but sub-linear, with improvements tapering as availability approaches the high-90\% range.}
        \label{fig:avai_data}
    \end{figure}

    Figure \ref{fig:avai_data} illustrates the relationship between the annual mean transmitted data and annual mean network availability. The plot demonstrates a clear upward trend, indicating that as the number of ground stations increases, both data transmission capacity and network availability improve significantly.

    The configuration with a single ground station shows the lowest mean transmitted data (just over 1,000 Gbits) and the lowest network availability (around 84\%). As additional ground stations are added, the transmitted data increases rapidly, reaching over 43,000 Gbits for a seven-station configuration, while network availability approaches 96\%. This trend demonstrates diminishing marginal returns on performance with each added OGS, suggesting the existence of an optimal number of stations for maximum network benefit.

\section{Discussion}

    The network availability, PDT and data throughput of four configurations of OGS network supporting a single LEO remote sensing satellite are compared. 
        
    This study presents a network performance estimation model similar to the one developed by DLR in the sense of leveraging pre-selected sites. However, the model presented in this paper goes beyond and distinguishes between large and small OGS capacities which is something that DLR's model does not include \cite{rattenbury_2024}. Our approach incorporates historical cloud and turbulence data with predefined OGS locations and characteristics (aperture size, power, data-rate) to distinguish between small and large OGS capacities.
    
    Network availability is defined here as having a cloud-free LOS to at least one OGS in the network. Although we must consider that certain OGS may be unavailable due to having no satellite in direct LOS or above the horizon of an OGS.
    
    The scenario hypothesizes the addition of a network of small-capacity OGS to complement a high-capacity OGS will translate to increased availability, data throughput and network efficiency. Preliminary results support this hypothesis, demonstrating that the small-capacity network offers significantly improved availability compared to the single large OGS network in a single location \cite{rattenbury_2024}. This higher availability in each satellite pass translates to a higher overall PDT for the small-capacity network. 

    Our study only includes single station availability for the OGS in Tenerife. We can see that the results obtained by this study differ from literature, which might be explained through further considerations to account for, such as pointing success, fine characterization of turbulence, equipment failure or inaccuracies, shorter time window used for the study, etc.
    
    In \cite{fuchs_2015}, five years of cloud data was evaluated cloud free availability of the OGS and provides results for three different network topologies (German, European, and intercontinental).
    
    The network availability is expected to improve with increasing the number of OGS and the geographical distribution of OGS. However, for our European network we have seen through our results that this is not a linear progression between performance and nodes in the network. However, there is an inflection point beyond which, despite a significant increase in the number of stations, the performance shows only a slight improvement. Considering Europe covers less than 2\% of global land mass, these results suggest distributing sites globally in other continents such as Africa, North America and Asia would increase this availability significantly. 

    The analysis followed in this paper can be applied to different fields to calculate expected performance. QKD, satellite communications, Earth Observation relay, and small satellites are examples of applications that can leverage from the analysis included in this study.

\subsection{Comparison with Previous Studies} 
    
    To validate our simulation model and understand its alignment with existing research, we compared our network availability results with those from other studies that utilized the same cloud data provider. Table~\ref{tab:network_availability} presents a comparative analysis between our model's availability estimates and those reported by Fuchs et al. \cite{fuchs_2015} and Rattenbury et al. \cite{cloud_cover_pearson} for various locations.
    
        \begin{table}[t]
        \centering
        \resizebox{\columnwidth}{!}{%
        \begin{tabular}{|l|c|c|}
        \hline
        \textbf{Location} & \textbf{Availability} $\eta_{\text{our}}$ & \textbf{Availability} $\eta_{\text{other}}$ \\
        \hline
        Tenerife          & 83.75\%                                   & 34.2\%                                      \\
        Nemea             & 31.58\%                                   & 60.0\%                                      \\
        Trauen            & 12.46\%                                   & 25.0\%                                      \\
        Madrid            & 25.91\%                                   & 63.7\%                                      \\
        Oberpfaffenhofen  & 19.22\%                                   & 31.2\%                                      \\
        \hline
        \end{tabular}%
        }
        \caption{Comparison of network availability estimates between our model ($\eta_{\text{our}}$) and previous studies ($\eta_{\text{other}}$) across different locations. Differences may be due to varying data collection periods and evolving cloud infrastructure.}
        \label{tab:network_availability}
        \end{table}

    \noindent Our study utilized cloud data from June 1, 2023, to June 1, 2024, while the other studies analyzed data from January 1, 2008, to January 1, 2013. The difference in time frames could contribute to the variations observed in network availability. For instance, our model predicts a significantly higher availability for Tenerife (83.75\%) compared to 34.2\% reported in previous work. Conversely, for locations like Nemea and Madrid, our availability estimates are lower than those found in the literature.
    
    \medskip
    \noindent Several additional factors might explain these discrepancies:
    \begin{itemize}
      \item \textbf{Temporal Variability.} Climate patterns and cloud cover can change over time. The different periods of data collection may reflect varying meteorological conditions, affecting availability estimates.
      \item \textbf{Modeling Assumptions.} Differences in how each study defines and calculates network availability can lead to varying results. Our model distinguishes between large and small OGS capacities, which may not be considered in other studies.
      \item \textbf{Spatial Resolution.} The cloud data's spatial resolution and the size of the bounding box used for averaging can influence the results. Our study averages cloud cover over a 20km$\times$20km area, which may differ from the methodologies used in other research.
    \end{itemize}
    
    \noindent Moreover, to achieve a more precise comparison, future work could involve running our model using cloud data from the same timeframe as the other studies (January 1, 2008, to January 1, 2013). This alignment would help isolate the impact of temporal variability in cloud cover and provide a clearer understanding of the discrepancies attributable to modeling assumptions and methodologies. By analyzing data from the same period, we can more accurately assess the differences between the models and enhance the validity of our comparative analysis.
    
    \subsection{Simulation Code Caveats}
        \label{chap: assumptions}
        
        We model the data downlink as a fixed value; however, in reality, factors such as turbulence, scintillation, and variations in the slant path influence both the effective data rates and the BER. 
        
        Also, when two OGS are within the line of sight of a satellite, the satellite must decide which OGS to communicate with. Currently, satellites prioritise the OGS with better visibility conditions, higher downlink rates, and closer proximity. Therefore, a decision-making logic is essential to determine which OGS to select for communication. 
        
        Additionally, our current code assumes a switching time of zero seconds between OGS, which is unrealistic. In practice, OGS tracking and locking systems require time to acquire the satellite, so understanding the acquisition time is critical for real-world applications. We additionally assume the satellite will communicate for the entire duration of a satellite pass. 
        
        The cloud data used in this study, sourced from EUMETSAT, has a temporal resolution of 15 minutes and a spatial resolution of around 3km. However, cloud cover in reality can change on a minute-by-minute basis, and LEO satellite passes often last for less than 15 minutes highlighting . The cloud cover over an OGS calculated by averaging the cloud cover values over a 20km x 20km bounding box, without understanding parameters such as cloud density or cloud base height, the bounding box in real life may be smaller or larger. This is to say that the specific cloud coverage threshold that results in a communication outage varies depending on conditions parameters.

    \subsection{Future Work}
    \label{chap: futurework}
        
        There are several potential areas for future work. The network simulation tool developed in this study can be tested under numerous scenarios. Incorporating multiple satellites, bidirectional links, data-relay systems, and mega-constellations within the simulation parameters could greatly enhance the insights the tool can provide. Further analysis could examine the feasibility of mobile terminals on vessels, vehicles, or trains.
        
        Exploring the logic behind OGS network and satellite tasking and management could enable users to envision the Concepts of Operations (CONOPS) of future proposed networks. This is increasingly relevant in today's ecosystem with the proliferation of Ground-Station-as-a-Service (GSaaS) products and autonomous and remote OGS operations. For example, innovations such as the Multi-User Acquisition and Tracking Sensor (MUATS) for optical communication, developed under ESA’s ARTES 4.0 programme, address existing limitations in satellite communication by providing a wide field of view and enabling simultaneous tracking of multiple OGS and satellites—significantly reducing the time needed to establish communication links between terminals compared to traditional space-to-ground FSO systems that can only connect with one partner at a time \cite{muats}. MUATS technology is laying the groundwork for autonomous space networks, with further developments such as the Autonomous Optical Terminal Detection Sensor aimed at facilitating the creation of fully autonomous and scalable optical satellite constellations.
        
        An OGS location optimization feature could significantly complement the network simulation's existing capabilities. Instead of restricting possible locations of the ground stations to a predefined set of sites around the globe, users could evaluate any point on Earth for the placement of an OGS.
        
        Meteorological categorizations can be enhanced through on-site meteorological instruments and sensors, or with mesoscale models that are more accurate than global-scale models averaged over a year. This could enable improved temporal and spatial resolutions of cloud coverage and turbulence statistics. Furthermore, additional sources of cloud data could be cross-checked. The current EUMETSAT CloudMask tool, limited to Europe and Africa, could be expanded to incorporate global cloud datasets from other data providers such as the National Oceanic and Atmospheric Administration (NOAA), enabling the simulation of worldwide OGS networks.
        
        Another area for improvement lies in the assumptions made during ground segment modeling, particularly the current use of a constant data rate. While a model has been developed to evaluate the link budget performance of ground stations, several important parameters—such as system noise and specific characteristics of the OGS—remain unaccounted for. Future models could improve upon this by calculating the data rate between the OGS and the receiver for specific use cases, considering link properties at given satellite and OGS positions. It is important to note that a link budget analysis has not been performed in this study, which would quantify how link performance changes over time. Conducting a detailed link budget analysis in future work would enable the simulation of how free-space losses and other factors dynamically affect link performance, offering a more accurate representation of real-world satellite communication.

    \subsection{Recommendations for Industry}
    \label{chap: recommendations}

        The number of FSO satellite missions is expected to rise significantly in the coming decade, driven by the advantages of FSO communications, technological advancements in the mitigation strategies and the success of recent demonstration missions. While FSO technology is not new, its implementation is still evolving. The NODES network exemplifies a global system with two high-capacity OGS in the progress of addressing challenges related to operations and performance. At the same time, a definitive use case and secure a strong customer base is yet to appear in the market.

        The availability of commercially produced OGS systems has expanded, signaling readiness for more satellite missions. With trends like GSaaS and remotely operated OGS systems, it is essential to provide quantitative justification for the cost and benefits of integrating FSO into existing infrastructures, such as the NODES network, or developing new ones.
        
        This study serves as a preliminary step toward creating a scalable and adaptable network simulation tool for various mission types and system requirements. Greater transparency and the sharing of mission performance data are needed from stakeholders to help cross-validate and improve network simulation models like ours. Collaborative opportunities exist with organisations such as EONN, NODES, and AOGS, which have access to existing databases, networks and satellite systems for testing.
        
        To facilitate wider adoption, it should be made easier for companies to integrate FSO technology into existing networks, with streamlined access to secure, high-data-rate solutions.
        
\section{Conclusions}

    The current state of FSO technology has been explored through an updates trade-off with RF satcomm, applications and global developments in the industrial ecosystem to identify the obstacles that stand in the way of the full realisation of optical space-to-ground communication networks. 
    
    Additionally, link performance and network availability trade- off studies are presented, comparing overall system performance of portable vs. large OGS networks in conjunction with a constellation of small LEO satellites. The feasibility of low-cost portable terminals as an alternative to larger high-capacity OGS systems.
    
    From our results, it can be concluded that increasing the number of OGS leads to higher network availability. However, the relationship between those two metrics is not linear and, after a threshold, adding more OGSs does not imply an increase in availability in the same percentage.
    
    Collaboration opportunities to support the provision of easy OGS network simulation modeling in the optical communications stakeholders, including governments and academic institutions, satellite operators, manufacturers, and communication service providers.
    
\section{Data Availability}

    The datasets and code supporting the findings of this study are publicly available in the GitHub repository at \url{https://github.com/connor-a-casey/fso-simulation}. This repository includes all necessary scripts, input data, and results to replicate and build upon the analyses presented in this paper. Additionally, any supplementary materials and documentation can be accessed within the repository to facilitate further research and collaboration.

\section{Acknowledgments}

    We would like to express our gratitude to the Space Generation Advisory Council (SGAC) and their Small Satellite Project Group for facilitating our collaboration on this project. Special thanks go to Simran Mardhini from Archangel Lightworks for her invaluable insights, support, and feedback throughout the research process. 

\section{Contributions}
C.C. carried out the data analysis, designed the simulation methodology, and implemented the entire codebase, refactoring all components to Python for ease of use and reproducibility. E.R. formulated the original concept as a literature review, which—together with C.C.’s modeling—evolved into the simulation study presented here. M.M. and K.W.V.T. contributed to the literature review and survey of prior work. K.W.V.T. also developed portions of the initial MATLAB codebase. I.P. contributed to software development. C.C. and E.R. wrote the majority of the manuscript, with input from all authors.

\section{Competing interests}
The authors declare no competing interests.

\bibliography{citations}

\appendix

\end{document}